\documentclass[twocolumn,showpacs,preprintnumbers,superscriptaddress,nofootinbib]{revtex4}
\usepackage{graphicx}

\newcommand{\mw}{m_W}
\newcommand{\mz}{m_Z}
\newcommand{\mt}{m_t}
\newcommand{\mb}{m_b}
\newcommand{\metwo}{m_{e}^2}
\newcommand{\mbtwo}{m_{b}^2}
\newcommand{\mttwo}{m_{t}^2}

\newcommand{\mhtwo}{m_{h^{0}}^2}
\newcommand{\mHtwo}{m_{H^{0}}^2}
\newcommand{\mhpmtwo}{m_{H^{\pm}}^2}
\newcommand{\mvtwo}{m_{V}^2}
\newcommand{\mwtwo}{m_{W}^2}
\newcommand{\mztwo}{m_{Z}^2}
\newcommand{\mstwo}{m_{S}^2}
\newcommand{\msitwo}{m_{S_i}^2}
\newcommand{\msjtwo}{m_{S_j}^2}
\newcommand{\msktwo}{m_{S_k}^2}

\newcommand{\ee}{e_e}
\newcommand{\enu}{e_{\nu}}
\newcommand{\eu}{e_u}
\newcommand{\ed}{e_d}

\newcommand{\gvel}{g_V^{eL}}
\newcommand{\gver}{g_V^{eR}}
\newcommand{\gvdl}{g_V^{dL}}
\newcommand{\gvdr}{g_V^{dR}}
\newcommand{\gvul}{g_V^{uL}}
\newcommand{\gvur}{g_V^{uR}}

\newcommand{\gznul}{g_Z^{\nu{L}}}
\newcommand{\gzel}{g_Z^{eL}}

\newcommand{\gzdl}{g_Z^{dL}}
\newcommand{\gzdr}{g_Z^{dR}}
\newcommand{\gzul}{g_Z^{uL}}
\newcommand{\gzur}{g_Z^{uR}}

\newcommand{\gpnul}{g_{\gamma}^{\nu{L}}}
\newcommand{\gpel}{g_{\gamma}^{eL}}

\newcommand{\gpdl}{g_{\gamma}^{dL}}
\newcommand{\gpdr}{g_{\gamma}^{dR}}
\newcommand{\gpul}{g_{\gamma}^{uL}}
\newcommand{\gpur}{g_{\gamma}^{uR}}

\newcommand{\yl}{y_{Htb}^L}
\newcommand{\yr}{y_{Htb}^R}
\newcommand{\ygl}{y_{Gtb}^L}
\newcommand{\ygr}{y_{Gtb}^R}

\newcommand{\aone}{{\cal{A}}_1}
\newcommand{\atwo}{{\cal{A}}_2}
\newcommand{\athree}{{\cal{A}}_3}
\newcommand{\afour}{{\cal{A}}_4}
\newcommand{\afive}{{\cal{A}}_5}
\newcommand{\asix}{{\cal{A}}_6}
\newcommand{\aseven}{{\cal{A}}_7}
\newcommand{\aeight}{{\cal{A}}_8}

\newcommand{\cw}{c_{W}}
\newcommand{\sw}{s_{W}}
\newcommand{\cba}{c_{\beta\alpha}}
\newcommand{\sba}{s_{\beta\alpha}}
\newcommand{\sinb}{\sin\beta}
\newcommand{\cosb}{\cos\beta}
\newcommand{\sat}{\sin\alpha_t}
\newcommand{\cat}{\cos\alpha_t}
\newcommand{\sab}{\sin\alpha_b}
\newcommand{\cab}{\cos\alpha_b}

\newcommand{\ghag}{g_{H^{-}AG^{+}}}
\newcommand{\ghhh}{g_{H^-h^{0}H^+}}
\newcommand{\ghHh}{g_{H^-H^{0}H^+}}
\newcommand{\ghhg}{g_{H^-h^{0}G^+}}
\newcommand{\ghHg}{g_{H^-H^{0}G^+}}
\newcommand{\gghg}{g_{G^-h^{0}G^+}}
\newcommand{\ggHg}{g_{G^-H^{0}G^+}}

\newcommand{\gw}{g_W}
\newcommand{\gpl}{g_{\gamma}^L}
\newcommand{\gzl}{g_{Z}^L}
\newcommand{\gpr}{g_{\gamma}^R}
\newcommand{\gzr}{g_{Z}^R}
\newcommand{\gvl}{g_{V}^L}
\newcommand{\ghl}{g_{H}^L}
\newcommand{\ggl}{g_{G}^L}
\newcommand{\gwl}{g_{W}^L}
\newcommand{\gvr}{g_{V}^R}
\newcommand{\ghr}{g_{H}^R}
\newcommand{\gwr}{g_{W}^R}
\newcommand{\ggr}{g_{G}^R}

\newcommand{\mi}{m_{\chi_i}}
\newcommand{\mj}{m_{\chi_j}}
\newcommand{\mk}{m_{\chi_k}}
\newcommand{\mitwo}{m_{\chi_i}^2}
\newcommand{\mjtwo}{m_{\chi_j}^2}
\newcommand{\mktwo}{m_{\chi_k}^2}
\newcommand{\msltwo}{m_{\tilde{l}}^2}
\newcommand{\mseltwo}{m_{\tilde{e}_L}^2}
\newcommand{\msnutwo}{m_{\tilde{\nu}}^2}

\newcommand{\slepton}{{\tilde{l}}}
\newcommand{\snu}{{\tilde{\nu}}}
\newcommand{\sel}{{\tilde{e}}_L}
\newcommand{\ser}{{\tilde{e}}_R}
\newcommand{\sul}{{\tilde{u}}_L}
\newcommand{\sdl}{{\tilde{d}}_L}
\newcommand{\stone}{{\tilde{t}}_{1}}
\newcommand{\sttwo}{{\tilde{t}}_{2}}
\newcommand{\sbone}{{\tilde{b}}_{1}}
\newcommand{\sbtwo}{{\tilde{b}}_{2}}
\newcommand{\sbl}{{\tilde{b}}_L}
\newcommand{\sbr}{{\tilde{b}}_R}
\newcommand{\stl}{{\tilde{t}}_L}
\newcommand{\str}{{\tilde{t}}_R}
\newcommand{\ghenu}{g_{H\sel\snu}}
\newcommand{\ghdu}{g_{H\sdl\sul}}
\newcommand{\ggenu}{g_{G\sel\snu}}
\newcommand{\ggdu}{g_{G\sdl\sul}}

\newcommand{\ghbonetone}{g_{H\sbone\stone}}
\newcommand{\ghbonettwo}{g_{H\sbone\sttwo}}
\newcommand{\ghbtwotone}{g_{H\sbtwo\stone}}
\newcommand{\ghbtwottwo}{g_{H\sbtwo\sttwo}}
\newcommand{\ggbonetone}{g_{G\sbone\stone}}
\newcommand{\ggbonettwo}{g_{G\sbone\sttwo}}
\newcommand{\ggbtwotone}{g_{G\sbtwo\stone}}
\newcommand{\ggbtwottwo}{g_{G\sbtwo\sttwo}}

\newcommand{\gchiie}{g_{{\chi_i}e\tilde{l}}}
\newcommand{\gchije}{g_{{\chi_j}e\tilde{l}}}
\newcommand{\gchike}{g_{{\chi_k}e\tilde{l}}}
\newcommand{\gchiiel}{g_{{\chi_i}e\tilde{l}}^L}

\newcommand{\gchijnur}{g_{{\chi_j}\nu\tilde{l}}^R}
\newcommand{\gchiiesel}{g_{{\chi_i}e\tilde{e}_L}^L}
\newcommand{\gchiinusnur}{g_{{\chi_i}\nu\tilde{\nu}}^R}
\newcommand{\gchiieser}{g_{{\chi_i}e\tilde{e}_L}^R}
\newcommand{\gchiiesnul}{g_{{\chi_i}e\tilde{\nu}}^L}
\newcommand{\gchiiesnur}{g_{{\chi_i}e\tilde{\nu}}^R}
\newcommand{\gchijeser}{g_{{\chi_j}e\tilde{e}_L}^R}
\newcommand{\gchijesel}{g_{{\chi_j}e\tilde{e}_L}^L}
\def\lsim{\mathrel{\raise.3ex\hbox{$<$\kern-.75em\lower1ex\hbox{$\sim$}}}}
\def\gsim{\mathrel{\raise.3ex\hbox{$>$\kern-.75em\lower1ex\hbox{$\sim$}}}}

\begin{document}

\preprint{CALT-68--2375}
\preprint{FERMILAB-Pub-02/046-T}
\preprint{hep-ph/0203270}

\title{Associated production of $H^{\pm}$ and $W^{\mp}$ 
in high-energy $e^+e^-$ collisions \\
in the Minimal Supersymmetric Standard Model}

\author{Heather E. Logan}
\email{logan@fnal.gov}
\affiliation{Theoretical Physics Department, Fermilab, PO Box 500, 
Batavia, Illinois 60510-0500, USA}

\author{Shufang Su}
\email{shufang@theory.caltech.edu}
\affiliation{California Institute of Technology, Pasadena, 
California 91125, USA}

\begin{abstract}
We study the associated production of the charged Higgs boson and 
$W^{\pm}$ gauge
boson in high energy $e^+e^-$ collisions in the Minimal Supersymmetric
Standard Model (MSSM).  This associated production, which first arises at
the one loop level, offers the possibility
of producing the charged Higgs boson at the $e^+e^-$ collider 
with mass more than half the center-of-mass energy, when the 
charged Higgs pair production is kinematically forbidden.
We present analytic and numerical results for the cross 
section for $e^+e^- \to W^+ H^-$ in the full MSSM, taking into account 
the previously uncalculated contributions from supersymmetric (SUSY) 
particles.
We find that the contributions of the SUSY particles enhance the 
cross section over most of SUSY parameter space, especially when the SUSY
particles are light, $\sim 200$ GeV.  
With favorable SUSY parameters, at small $\tan\beta$, 
this process can yield more than
ten $W^{\pm}H^{\mp}$ events for $m_{H^{\pm}} \lsim 350$ GeV in 500 fb$^{-1}$
at a 500 GeV $e^+e^-$ collider, or $m_{H^{\pm}} \lsim 600$ GeV 
in 1000 fb$^{-1}$
at a 1000 GeV collider.  80\% left-handed polarization of the $e^-$ beam
improves these reaches to $m_{H^{\pm}} \lsim 375$ GeV and 
$m_{H^{\pm}} \lsim 670$ GeV, respectively.
\end{abstract}


\maketitle


\section{\label{sec:intro}Introduction}

Discovery of the heavy Higgs bosons $H^0$, $A^0$ and $H^{\pm}$
of the Minimal Supersymmetric Standard Model (MSSM)\footnote{For a 
pedagogical introduction to the MSSM Higgs sector, see 
Refs.~\cite{HHG,GunionHaber}.}
poses a special challenge at future colliders.
Run~II of the Fermilab Tevatron, now in progress, has a limited 
reach for the neutral heavy MSSM Higgs bosons.  It can detect 
the CP-even and CP-odd neutral Higgs bosons $H^0$ and $A^0$
if their masses are below about 150 GeV and $\tan\beta$ (the ratio of the
two Higgs vacuum expectation values) is large, so that the 
couplings of $H^0$ and $A^0$ to $b$ quarks are
enhanced~\cite{Dai:1996rn,Balazs:1998nt,HiggsatRunII,CMWpheno}.
For the charged Higgs boson $H^{\pm}$, 
no sensitivity is expected via direct production 
at the Tevatron unless QCD and SUSY effects conspire to 
enhance the cross section \cite{BelyaevHpm}; 
however, $H^{\pm}$ can be discovered in top quark decays
for $m_{H^{\pm}} \lsim m_t$ and large $\tan\beta$ 
\cite{HiggsatRunII,GuchaitHpm}.
The CERN Large Hadron Collider (LHC) has a much greater reach for heavy
MSSM Higgs boson discovery at moderate to large values of $\tan\beta$.  
$H^0$ and $A^0$ can be discovered with decays to $\tau$ pairs for
$\tan\beta \gsim 10$ for a CP-odd Higgs mass of $m_{A^0} = 250$ GeV 
($\tan\beta \gsim 17$ for $m_{A^0} = 500$ GeV)~\cite{AtlasTDR,CMS}. 
The charged MSSM Higgs boson $H^{\pm}$ 
can be discovered in $gb \to t H^+$ with $H^+ \to \tau \nu$
for virtually the same $\tan\beta$ values~\cite{AtlasH+,CMS}.
The absence of a Higgs boson discovery at the CERN LEP-2 experiments
implies that $0.5 < \tan\beta < 2.4$ and $m_{A^0} < 91.9$ GeV
are excluded at 95\% confidence level \cite{LEP2}.
This leaves a wedge-shaped region of parameter space at moderate $\tan\beta$ 
in which the heavy MSSM Higgs bosons will not be discovered at the 
LHC.
At a future high energy $e^+e^-$ linear collider (LC), the heavy Higgs
bosons will be produced in pairs, if it is kinematically allowed.
The dominant production modes are $e^+e^- \to H^0 A^0$ and 
$e^+e^- \to H^+H^-$; for experimental studies see 
Refs.~\cite{Kiiskinen,Andreazza}.
These production modes are kinematically allowed only if 
$m_{H^0} + m_{A^0} < \sqrt{s}$ and $2 m_{H^{\pm}} < \sqrt{s}$, respectively,
where $\sqrt{s}$ is the center-of-mass energy 
of the LC.\footnote{The pair production of scalars is $p$-wave suppressed
near threshold, so in practice the Higgs mass reach in these modes is likely
to be somewhat below the kinematic limit.}
At large $m_{A^0}$, $m_{A^0} \simeq m_{H^0} \simeq m_{H^{\pm}}$
up to mass splittings of order $m_Z^2/m_{A^0}$, so that the pair-production
modes are kinematically allowed only if $m_{A^0} \lsim 0.5 \sqrt{s}$.
In particular, the pair-production modes are limited to 
$m_{A^0} \lsim 250$ GeV ($m_{A^0} \lsim 500$ GeV) at a LC with
$\sqrt{s} = 500$ GeV ($\sqrt{s} = 1000$ GeV).

By contrast, the discovery of the light MSSM Higgs boson $h^0$, 
which typically has properties similar to those of the SM Higgs boson,
is much more certain.  
At Run~II of the Tevatron, discovery of $h^0$ at the $5 \sigma$ level 
is possible for 
$m_{h^0} \lsim 120$ GeV if 15 fb$^{-1}$ of integrated luminosity 
can be collected and the detectors perform as expected~\cite{HiggsatRunII}.  
This covers a large fraction of the MSSM parameter space left unexcluded
after LEP-2 \cite{LEP2}.
At the LHC, discovery of $h^0$ is virtually guaranteed over all of the
MSSM parameter space \cite{AtlasTDR}.  Enough $h^0$ events are expected in
a number of different production and decay modes to allow the measurement
of various combinations of $h^0$ partial widths with precisions on the order
of 15\%~\cite{Zeppenfeld}.
At the LC, $h^0$ will be copiously produced.
The promise of the LC for making precision measurements of the couplings
of $h^0$ at the few-percent level has been well 
documented~\cite{hLCmeas,TeslaTDR,Orangebook}.  
If the MSSM Higgs sector 
is not too far into the decoupling limit~\cite{decoupling}, 
in which the heavy Higgs bosons become increasingly heavy and 
the couplings of $h^0$ to SM particles approach their SM values
(so that $h^0$ behaves like the SM Higgs boson),
the LC measurements of $h^0$ couplings can be used to distinguish 
$h^0$ from the SM Higgs boson and to extract MSSM
parameters~\cite{CHLM}.  One finds $\geq 2 \sigma$ deviations from the
SM typically for $m_{A^0} \lsim 600$ GeV~\cite{hLCmeas,CHLM}.\footnote{A subset
of the $h^0$ couplings can also be measured with similar precision
at a LC operating as a $\gamma\gamma$ collider, leading again to 
$\geq 2 \sigma$ deviations from the
SM typically for $m_{A^0} \lsim 600$ GeV~\cite{Velascogg}.}

While in the decoupling limit
the couplings of $h^0$ become increasingly insensitive to the MSSM parameters,
the couplings of the heavy Higgs bosons exhibit no such decoupling; 
rather, they are always sensitive to the MSSM parameters.  Further,
the measurements of the masses and couplings of the heavy Higgs bosons
in addition to those of $h^0$ allow one to place orthogonal constraints 
on the parameters of the MSSM Higgs sector.
Thus, measurements of the properties of the heavy MSSM Higgs bosons 
are very valuable, especially in the decoupling limit.

In this paper we consider the production of one of the heavy Higgs bosons 
alone or in association with lighter SM particles at the LC.
While the cross sections for such production modes are typically very
small, they offer the possibility of extending the reach of the LC
to higher values of $m_{A^0,H^0,H^{\pm}} \gsim 0.5 \sqrt{s}$.  
Single heavy Higgs boson production has been studied in the context of
the MSSM or a general two Higgs doublet model (2HDM)
in a number of processes.
The following final states have been considered in $e^+e^-$ collisions:
$ZH^0$ and $h^0 A^0$~\cite{FDHiggsprod}; 
$\nu \bar \nu H^0$ and $e^+e^- H^0$ (see, {\it e.g.}, Ref.~\cite{Orangebook});
$b \bar b H^0$, $b \bar b A^0$, $\tau^- \nu H^+$
and $\bar t b H^+$~\cite{bbHAHpm,bbHA,bbHpm,KanemuraReview};
$Z A^0$~\cite{eeZA2HDM,eeZAMSSM}; $\gamma A^0$~\cite{gammaA,eeZA2HDM,FGL};
and $W^+H^-$~\cite{eeWH2HDM,KanemuraeeHW,SHZhu}.
If the $e^+e^-$ LC is converted into a photon collider through Compton 
backscattering of intense laser beams, single heavy Higgs bosons can
be produced via $\gamma\gamma$ collisions.  In particular,
$H^0$ and $A^0$ can be produced in the 
$s$-channel~\cite{gamgamAsner,gamgamMuhlleitner,gamgamGunionHaber};
the final states $\tau^- \nu H^+$ and $\bar t b H^+$~\cite{CPYuan}
and $W^+H^-$~\cite{ggHW} are also accessible.
Finally, in $e^- \gamma$ collisions one can produce
$\nu H^-$~\cite{Kanemuraegamma}.

In this paper we compute the cross section for $e^+e^- \to W^+H^-$ in
the full MSSM.  We present analytic formulae for the matrix elements
and numerical results for the cross section as a function of the MSSM
parameters.
We confirm the calculation of $e^+e^- \to W^+H^-$ in the non-supersymmetric
2HDM performed in Refs.~\cite{eeWH2HDM,KanemuraeeHW,SHZhu}.\footnote{A 
significant contribution to $e^+e^- \to W^+H^-$
comes from the loop-induced $W^+H^-Z$ and $W^+H^-\gamma$ vertices.
The contributions to these vertices from top/bottom quark loops
\cite{PHI} and gauge and Higgs boson loops \cite{KanemuraWHZ} 
have been computed in the 2HDM.
The full MSSM contributions to these vertices have been computed in 
Ref.~\cite{Iruleguithesis}; 
the contributions from top/bottom squark loops are also given 
in Ref.~\cite{PHI}.  The loop-induced $W^+H^-Z$ vertex
has also been computed numerically in the full MSSM in Ref.~\cite{WHZMSSM}.}
We find that the contributions of the SUSY particles enhance the cross
section over most of SUSY parameter space, especially when the SUSY 
particles are light, with masses of order 200 GeV.
The largest contributions to the cross section from the SUSY sector 
come from diagrams involving charginos/neutralinos and top/bottom squarks.
We also find that left-handed polarization of the $e^-$ beams leads
to an order 50\% enhancement of the cross section.
Throughout this paper we present
cross sections for the single process $e^+e^- \to W^+H^-$; these numbers 
should be doubled to find the combined cross sections
for this process plus its charge conjugate.

This paper is organized as follows.  In Sec.~\ref{sec:formalism} 
we introduce our formalism for the matrix elements and cross section.
In Sec.~\ref{sec:renorm} we display the relevant SUSY diagrams
and review the renormalization procedure.
In Sec.~\ref{sec:numerics} we present our numerical results.
In Sec.~\ref{sec:comparison} we review the literature on other single
heavy Higgs boson production processes and compare their reach to 
that of $e^+e^- \to W^+H^-$.
Sec.~\ref{sec:conclusions} is reserved for our conclusions.
The matrix elements for the 2HDM and SUSY diagrams are collected in
the appendix.

\section{\label{sec:formalism}Formalism}

Following the notation of Ref.~\cite{eeWH2HDM},
the matrix element $\cal{M}$ for $e^+e^- \to W^+H^-$ 
can be decomposed into six independent 
matrix elements ${\cal{A}}_i$
and their corresponding coefficients ${\cal{M}}_i$:
\begin{equation}
{\cal{M}}=\sum_{i=1}^{6}{\cal{M}}_i{\cal{A}}_i.
\label{eq:calM}
\end{equation}

The six matrix elements are defined as,
\begin{eqnarray}
{\cal{A}}_1&=&\bar{v}(p_2)\not\epsilon^*(k_1)
	\frac{1+\gamma_5}{2}u(p_1)\nonumber \\
{\cal{A}}_2&=&\bar{v}(p_2)\not\epsilon^*(k_1)
	\frac{1-\gamma_5}{2}u(p_1)\nonumber \\
{\cal{A}}_3&=&\bar{v}(p_2)\not{k}_1\frac{1+\gamma_5}{2}u(p_1)
	(p_1\cdot\epsilon^*(k_1))\nonumber \\
{\cal{A}}_4&=&\bar{v}(p_2)\not{k}_1\frac{1-\gamma_5}{2}u(p_1)
	(p_1\cdot\epsilon^*(k_1))\nonumber \\
{\cal{A}}_5&=&\bar{v}(p_2)\not{k}_1\frac{1+\gamma_5}{2}u(p_1)
	(p_2\cdot\epsilon^*(k_1))\nonumber \\
{\cal{A}}_6&=&\bar{v}(p_2)\not{k}_1\frac{1-\gamma_5}{2}u(p_1)
	(p_2\cdot\epsilon^*(k_1)),
\label{eq:meleA}
\end{eqnarray}
where $\epsilon^*$ is the polarization vector of the $W^+$ boson,
$p_1$ and $p_2$ are the incoming momenta of the initial $e^-$ and $e^+$,
respectively, and $k_1$ is the outgoing momentum of the $W^+$.

For convenience we define two additional matrix elements,
$\aseven$ and $\aeight$:
\begin{eqnarray}
{\cal{A}}_7&=&i\varepsilon^{\alpha\mu\beta\gamma}\bar{v}(p_2)\gamma_{\mu}
\frac{1+\gamma_5}{2}u(p_1)\epsilon_{\alpha}^*k_{1\beta}k_{2\gamma}\nonumber \\
{\cal{A}}_8&=&i\varepsilon^{\alpha\mu\beta\gamma}\bar{v}(p_2)\gamma_{\mu}
\frac{1-\gamma_5}{2}u(p_1)\epsilon_{\alpha}^*k_{1\beta}k_{2\gamma},
\label{eq:a7a8}
\end{eqnarray}
where $k_2 = p_1 + p_2 - k_1$ is the outgoing momentum of the $H^-$ and
$\varepsilon^{\alpha\mu\beta\gamma}$ is the totally antisymmetric tensor,
with $\varepsilon^{0123} = 1$.
$\aseven$ and $\aeight$ can be expressed in terms of $\aone$--$\asix$
as follows:
\begin{eqnarray}
\aseven&=&\aone(t-u)/2+\athree-\afive \nonumber \\
\aeight&=&\atwo(u-t)/2-\afour+\asix,
\label{eq:a7a8toA}
\end{eqnarray}
where $t = (p_1 - k_1)^2$ and $u = (p_1 - k_2)^2$.

From the form of these matrix elements, one immediately sees that
the cross sections for like polarizations of $e^+$ and $e^-$ are zero:
\begin{equation}
	\frac{d \sigma (e^+_R e^-_R \to W^+ H^-)}{d \cos\theta}
	= \frac{d \sigma (e^+_L e^-_L \to W^+ H^-)}{d \cos\theta}
	= 0.
\end{equation}
For unlike polarizations of $e^+$ and $e^-$, the cross sections are given
in terms of the six basic matrix elements by:
\begin{widetext}
\begin{eqnarray}
	\frac{d \sigma (e^+_R e^-_L \to W^+ H^-)}{d \cos\theta} &=& 
	\frac{\kappa}{32 \pi s} \left[ 2s |\mathcal{M}_2|^2 
	- \frac{(m_{H^{\pm}}^2 m_W^2 - tu)}{4m_W^2} \left\{
	(m_W^2 - t)^2 |\mathcal{M}_4|^2
	+ (m_W^2 - u)^2 |\mathcal{M}_6|^2 \frac{}{} \right. \right.
	\nonumber \\
	&& \hskip-3.6cm \left. \left. \frac{}{}
	+ 4(m_W^2 - t) {\rm Re}[\mathcal{M}_2 \mathcal{M}_4^*]
	+ 4(m_W^2 - u) {\rm Re}[\mathcal{M}_2 \mathcal{M}_6^*]
	+ 2(tu - m_W^2 s - m_W^2 m_{H^{\pm}}^2) 
	{\rm Re}[\mathcal{M}_4 \mathcal{M}_6^*] + 4 |\mathcal{M}_2|^2
	\right\} \right],
\end{eqnarray}
and
\begin{eqnarray}
	\frac{d \sigma (e^+_L e^-_R \to W^+ H^-)}{d \cos\theta} &=& 
	\frac{\kappa}{32 \pi s} \left[ 2s |\mathcal{M}_1|^2 
	- \frac{(m_{H^{\pm}}^2 m_W^2 - tu)}{4m_W^2} \left\{
	(m_W^2 - t)^2 |\mathcal{M}_3|^2
	+ (m_W^2 - u)^2 |\mathcal{M}_5|^2 \frac{}{} \right. \right.
	\nonumber \\
	&& \hskip-3.6cm \left. \left. \frac{}{}
	+ 4(m_W^2 - t) {\rm Re}[\mathcal{M}_1 \mathcal{M}_3^*]
	+ 4(m_W^2 - u) {\rm Re}[\mathcal{M}_1 \mathcal{M}_5^*]
	+ 2(tu - m_W^2 s - m_W^2 m_{H^{\pm}}^2) 
	{\rm Re}[\mathcal{M}_3 \mathcal{M}_5^*] + 4 |\mathcal{M}_1|^2
	\right\} \right].
\end{eqnarray}
\end{widetext}
Here $s = (p_1 + p_2)^2$ and $\kappa = 2 |\vec k_1| / \sqrt{s}$ is given by:
\begin{equation}
	\kappa^2 = \left[1 - \frac{(m_{H^{\pm}} + m_W)^2}{s} \right]
		\left[1 - \frac{(m_{H^{\pm}} - m_W)^2}{s} \right].
\end{equation}
The unpolarized cross section is obtained by averaging over the four
possible initial combinations of $e^+e^-$ polarizations.

\section{\label{sec:renorm}Renormalization}

\begin{figure}
\resizebox{8.5cm}{!}{\includegraphics*[43,296][541,739]{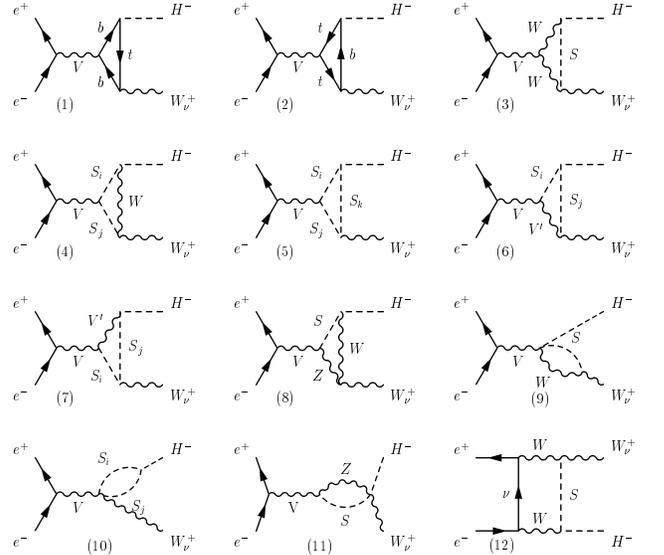}}
\caption{Feynman diagrams for the 2HDM contributions to $e^+e^- \to
W^+H^-$.  Here $S$, $S_{i,j,k}$ denote Higgs and Goldstone bosons,
$V = \gamma$, $Z$, and $V^{\prime}= Z$, $W^{\pm}$.}
\label{fig:2HDMdiagrams}
\end{figure}

\begin{figure}
\resizebox{8.5cm}{!}{\includegraphics*[43,296][541,739]{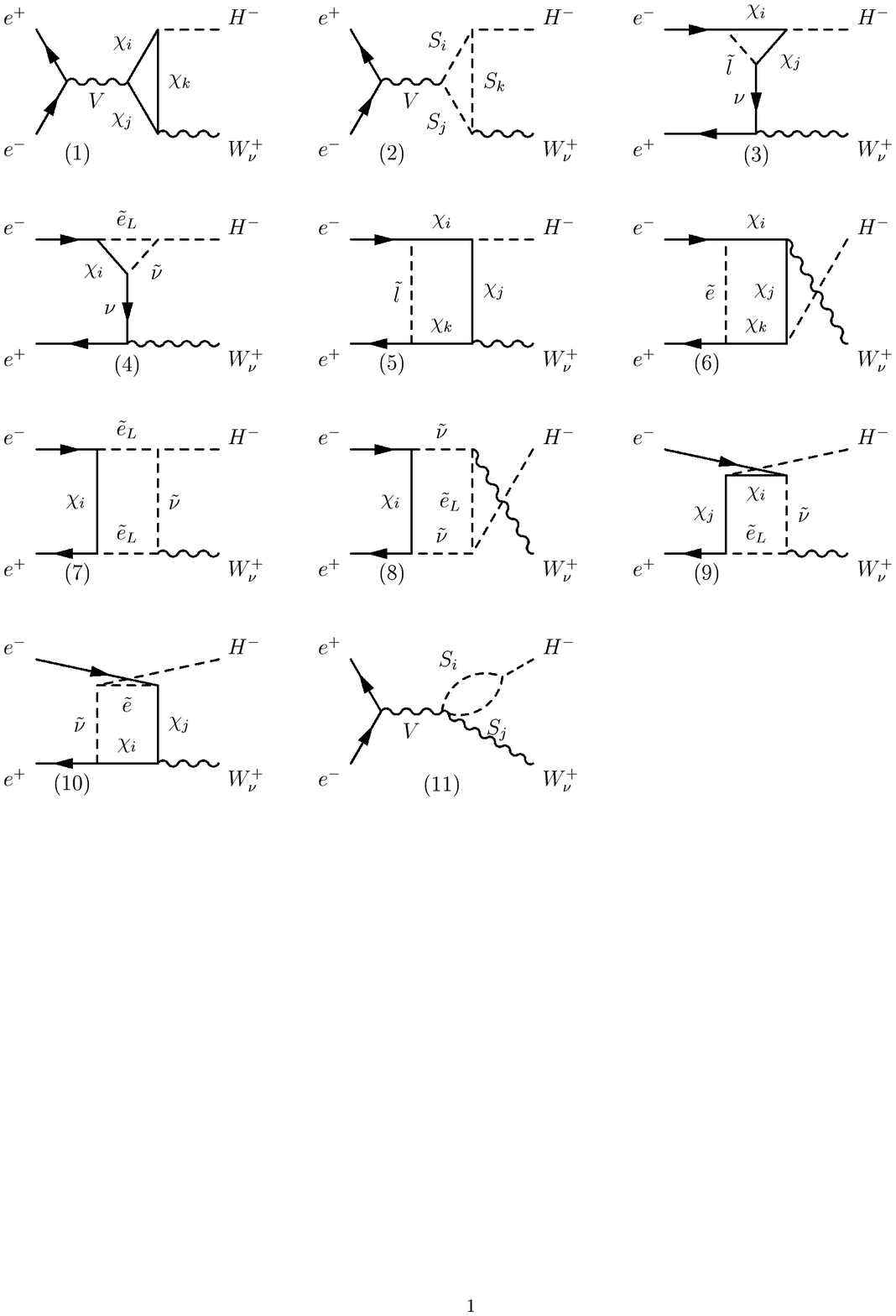}}
\caption{Feynman diagrams for the SUSY contributions to $e^+e^- \to
W^+H^-$.  Here $\chi_{i,j,k}$ denote charginos and neutralinos,
$S_{i,j,k}$ denote squarks and sleptons, and $V = \gamma$, $Z$.}
\label{fig:susydiagrams}
\end{figure}

\begin{figure}
\resizebox{8.5cm}{!}{\includegraphics*[40,533][540,735]{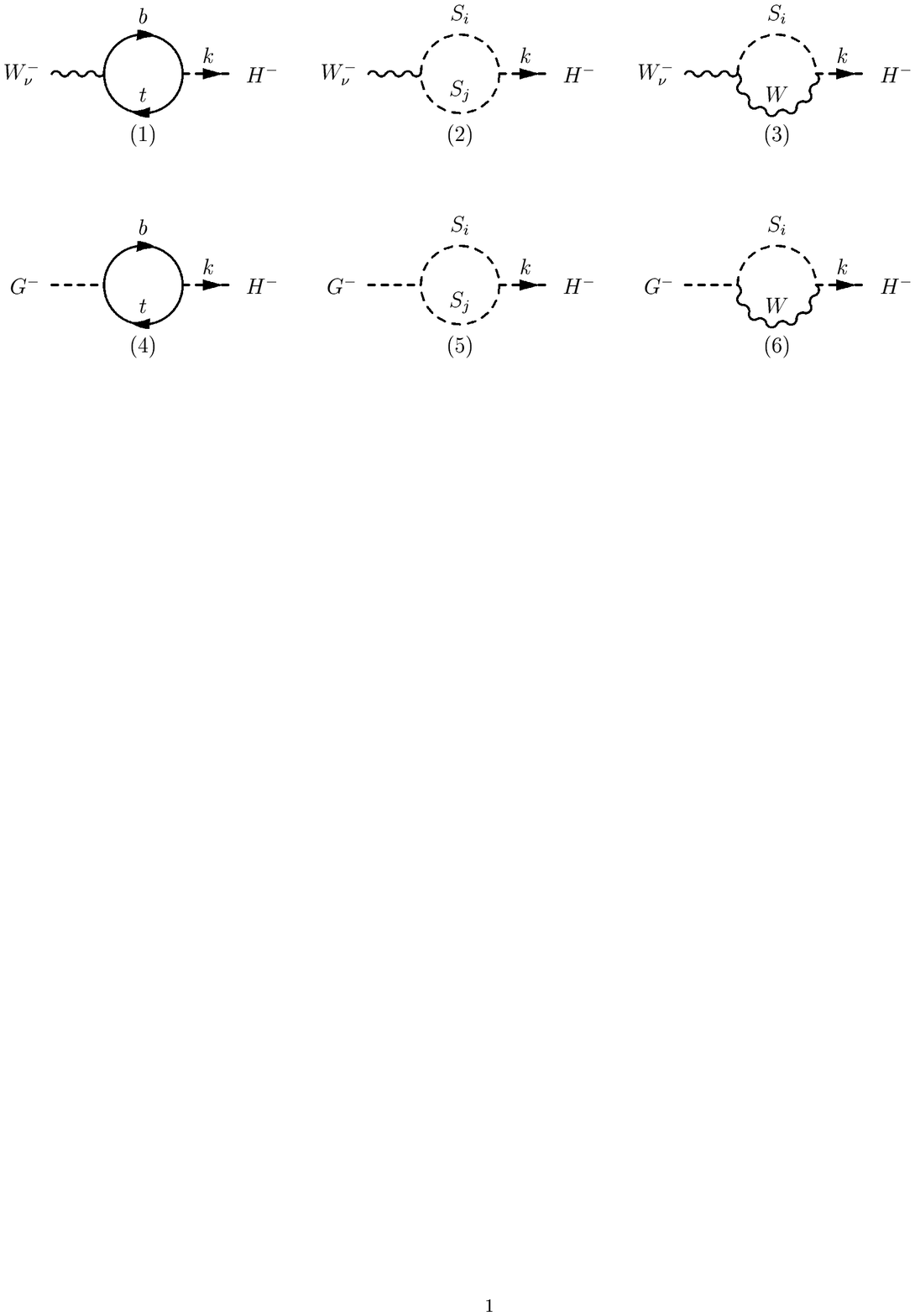}}
\caption{Feynman diagrams for the 2HDM contributions to $W^-H^-$ and
$G^-H^-$ mixing self-energies.  Here 
$S_i=h^0,H^0$ and $S_j=H^{\pm}, G^{\pm}$. }
\label{fig:2HDMselfenergy}
\end{figure}

\begin{figure}
\resizebox{8.5cm}{!}{\includegraphics*[40,533][366,735]{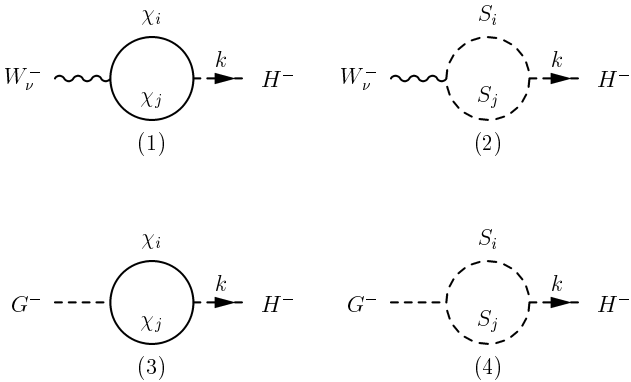}}
\caption{Feynman diagrams for the SUSY contributions to $W^-H^-$ and
$G^-H^-$ mixing self-energies.  Here $\chi_{i,j}$ 
denote charginos and neutralinos
and $S_{i,j}$ denote squarks and sleptons.}
\label{fig:susyselfenergy}
\end{figure}

\begin{figure}
\resizebox{8.5cm}{!}{\includegraphics*[40,533][366,735]{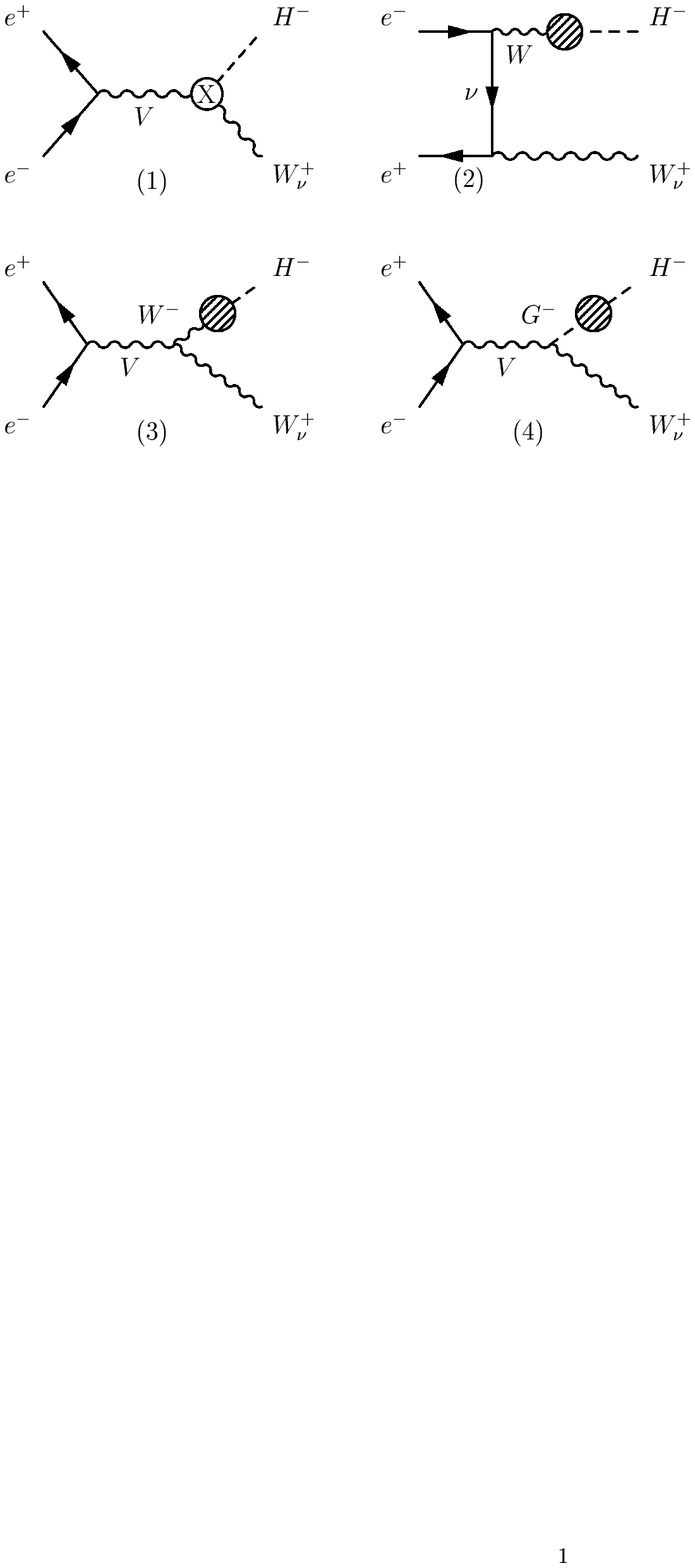}}
\caption{Feynman diagrams for the contributions to $e^+e^- \to W^+H^-$
through the counterterms and the $W^-H^-$ and $G^-H^-$ mixing self-energies.
The ``X'' in diagram 1 denotes the counterterm for the
$\gamma{W}^+H^-$ and $ZW^+H^-$ vertex, and the blob in diagrams 
2-4 denotes the renormalized $W^-H^-$ and $G^-H^-$ mixing.  }
\label{fig:WHGH}
\end{figure}

The diagrams contributing to $e^+e^- \to W^+H^-$ in the 2HDM are 
shown in Fig.~\ref{fig:2HDMdiagrams} \cite{eeWH2HDM}.
We show the additional SUSY diagrams in Fig.~\ref{fig:susydiagrams}.
There are also $W^-H^-$ and $G^-H^-$ mixing self-energy diagrams that
involve SM fermions, gauge and Higgs bosons (Fig.~\ref{fig:2HDMselfenergy}), 
and SUSY particles (Fig.~\ref{fig:susyselfenergy}).
These mixing self-energies, together with the counterterms, 
contribute to $e^+e^- \to W^+H^-$ via the diagrams in Fig.~\ref{fig:WHGH}.
        
\begin{figure}
\resizebox{8.5cm}{!}{\includegraphics*[45,530][538,738]{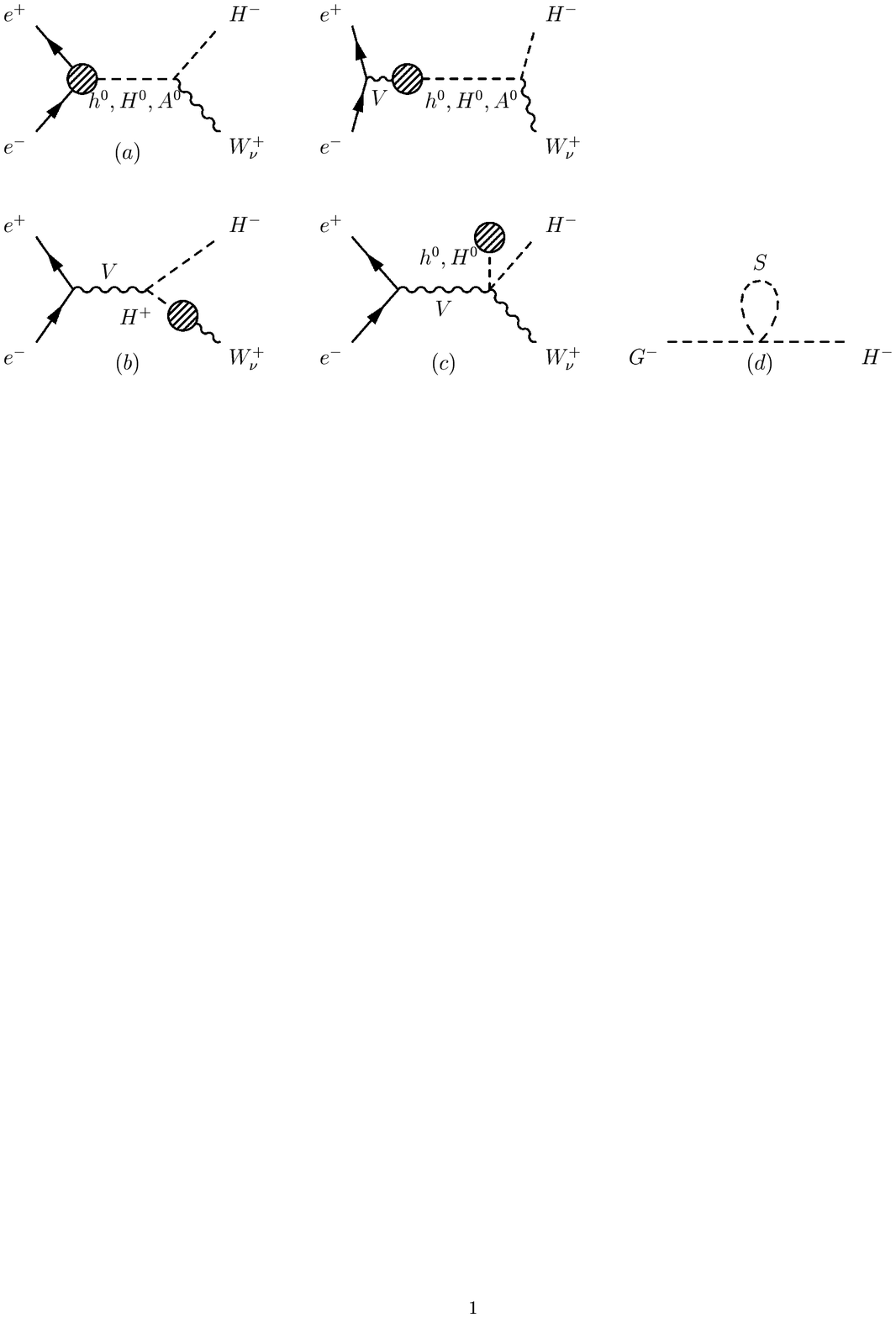}}
\caption{Contributions to $e^+e^- \to W^+H^-$ that we neglect.}
\label{fig:null}
\end{figure}

We neglect all diagrams that are proportional to the electron
Yukawa coupling.
We also neglect the diagrams shown in Fig.~\ref{fig:null}:
\begin{itemize}
\item[(a)] These two diagrams are
proportional to $m_e$ either through
the electron Yukawa coupling or through the factor of $m_e$ obtained via
the equation of motion of the incoming electron. 
In particular, the vector boson--Higgs mixing is proportional 
to $k^V_{\mu}=p_{1\mu}+p_{2\mu}$,
which gives $m_e$ when acting on the $V_\mu\bar{e}\gamma^\mu{e}$ vertex.
\item[(b)] This diagram is zero because
the $W^-H^-$ mixing self-energy is proportional to $k_{1\nu}$, and 
$k_{1}\cdot\epsilon^{*}=0$.
\item[(c)] We will set the renormalized tadpoles to zero below, so that 
this diagram does not contribute.
(Note that 
the $A^0$ and $G^0$ tadpoles are zero automatically due to CP conservation.)
\item[(d)] This diagram is purely real and 
is canceled by the $G^-H^-$ mixing counterterm, as discussed below.
\end{itemize}
       
We evaluate the one-loop process $e^+e^- \to W^+H^-$
in the 't Hooft--Feynman gauge using dimensional regularization.
Using dimensional reduction yields the same result for this process.
Diagrams that contain closed loops of chiral fermions may give
rise to inconsistencies in the treatment of $\gamma^5$ 
in na\"{\i}ve dimensional 
regularization~\cite{Breitenlohner,NDRinconsistencies}.
We have checked that this does not happen in the one-loop 
$e^+e^- \to W^+H^-$ diagrams by computing 
the terms involving $\gamma^5$ in the 
diagrams with closed fermion loops using the 't Hooft-Veltman 
rules~\cite{tHooftVeltman,Breitenlohner}.
It was shown in Ref.~\cite{Breitenlohner}
that this prescription yields a consistent formulation of dimensional 
regularization even if $\gamma^5$ couplings are present.  
(For a pedagogical discussion see Ref.~\cite{BurasWeisz}.)
We find that in this process the 't Hooft-Veltman rules yield the 
same result as na\"{\i}ve dimensional regularization, so that 
no inconsistencies arise.

We follow the on-shell renormalization formalism developed
by Dabelstein~\cite{Dabelstein} for the MSSM gauge and Higgs sectors.
The one-particle-irreducible (1PI) tadpole diagrams 
for $h^0$ and $H^0$ are given 
by $-iT_h$ and $-iT_H$, respectively.  Including the tadpole counterterms,
the renormalized tadpoles are given by:
\begin{equation}
	\hat T_h = T_h + \delta t_h, \ \ \ \hat T_H = T_H + \delta t_H.
\label{eq:tadpoleCTs}
\end{equation}

The 1PI two-point function for $W^+H^+$ mixing is defined as 
$-i k_{\mu} \Sigma_{W^+H^+}(k^2)$, where $k$ is the incoming momentum
of the $W^+_{\mu}$, and $H^+$ is outgoing.\footnote{We use the convention
$\mathcal{D}_{\mu} = \partial_{\mu} + i e A_{\mu}$.  Ref.~\cite{eeWH2HDM}
uses the opposite convention, so our matrix elements should differ
from theirs by a sign.}
The conjugate
two-point function, with $W^-$ incoming and $H^-$ outgoing, is given
by $+i k_{\mu} \Sigma_{W^+H^+}(k^2)$, where again $k$ is the incoming 
momentum of the $W^-_{\mu}$.
The renormalized two-point function for $W^+H^+$ mixing 
is obtained by adding the counterterm:
\begin{eqnarray}
	\hat \Sigma_{W^+H^+}(k^2)
	&=& \Sigma_{W^+H^+}(k^2) \nonumber \\
	&& \hskip-2.5cm - m_W \sin\beta \cos\beta
	\left(\delta Z_{H_1} - \delta Z_{H_2} 
	- \frac{\delta v_1}{v_1} + \frac{\delta v_2}{v_2} \right),
\label{eq:WHct}
\end{eqnarray}
where $\delta Z_{H_1}$ and $\delta Z_{H_2}$ renormalize the two Higgs
doublet fields $H_1$ and $H_2$ and $\delta v_1$ and $\delta v_2$ renormalize
the two Higgs vacuum expectation values~\cite{Dabelstein}:
\begin{equation}
	H_i \to Z_{H_i}^{1/2} H_i; \qquad
	v_i \to Z_{H_i}^{1/2}(v_i - \delta v_i),
\end{equation}
and $Z_{H_i} = 1 + \delta Z_{H_i}$.

The 1PI two-point function for $G^+H^+$ mixing is defined as
$+i \Sigma_{G^+H^+}(k^2)$, where $k$ is the incoming momentum of
the $G^+$, and $H^+$ is outgoing.  The conjugate two-point function,
with $G^-$ incoming and $H^-$ outgoing, is the same.
The renormalized two-point function for $G^+H^+$ mixing,
$\hat \Sigma_{G^+H^+}(k^2)$,
is fixed in terms of $\hat \Sigma_{W^+H^+}(k^2)$
due to the Slavnov-Taylor
identity (see, {\it e.g.}, Refs.~\cite{SlavnovTaylor,SlavnovTaylor2} 
for details):
\begin{equation}
	k^2 \hat \Sigma_{W^+H^+}(k^2) - m_W \hat \Sigma_{G^+H^+}(k^2) = 0.
\label{eq:slavnovtaylor}
\end{equation}

Finally, there are two vertex counterterms for
$\gamma_{\mu} W^+_{\nu} H^-$ and $Z_{\mu} W^+_{\nu} H^-$ (all particles
outgoing).  These are given by
$-iem_W \sin\beta \cos \beta \, \delta c \, g_{\mu\nu}$
and 
$+ies_W m_Z \sin\beta \cos \beta \, \delta c \, g_{\mu\nu}$,
respectively.  Here $\delta c \equiv (\delta Z_{H_1} - \delta Z_{H_2}
- \delta v_1/v_1 + \delta v_2/v_2)$ and
$s_W$ denotes the sine of the weak mixing angle.
Note that $\delta c$ is fixed in terms of the $W^+H^+$ 
mixing counterterm, Eq.~\ref{eq:WHct}.

We need impose only the following two 
renormalization conditions.\footnote{Because the process $e^+e^- \to W^+H^-$
is zero at tree level, the renormalization procedure is greatly simplified
and many renormalization conditions, such as that for $\tan\beta$, 
need not be imposed.}  
First, the renormalized tadpoles in Eq.~\ref{eq:tadpoleCTs} are set to zero.
Second, the real part of the renormalized $W^+H^+$ mixing is set to zero 
when $H^+$ is on mass shell:
\begin{equation}
	{\rm Re} \, \hat \Sigma_{W^+H^+}(m_{H^{\pm}}^2) = 0.
\label{eq:WHrenor}
\end{equation}
This fixes the following combination of counterterms:
\begin{eqnarray}
	 && m_W \sin\beta \cos\beta
	\left(\delta Z_{H_1} - \delta Z_{H_2} 
	- \frac{\delta v_1}{v_1} + \frac{\delta v_2}{v_2} \right) 
	\nonumber \\
	&& \hskip1cm = {\rm Re} \, \Sigma_{W^+H^+}(m_{H^{\pm}}^2),
\label{eq:counter}
\end{eqnarray}
which appears in the
$\gamma W^+H^-$ and $Z W^+H^-$ vertex counterterms.  
Applying the
Slavnov-Taylor identity (Eq.~\ref{eq:slavnovtaylor}) at $k^2 = m_{H^{\pm}}^2$,
this condition also
fixes ${\rm Re} \, \hat \Sigma_{G^+H^+}(m_{H^{\pm}}^2) = 0$.
In addition, since the counterterms are purely real, we have 
from Eq.~\ref{eq:slavnovtaylor}:
\begin{equation}
	{\rm Im} \, \Sigma_{G^+H^+}(m_{H^{\pm}}^2) 
	= \frac{m^2_{H^{\pm}}}{m_W} {\rm Im} \, \Sigma_{W^+H^+}(m_{H^{\pm}}^2),
\label{eq:slavnovtaylor2}
\end{equation} 
so that diagrams 4-6 of Fig.~\ref{fig:2HDMselfenergy},
diagrams 3 and 4 of Fig.~\ref{fig:susyselfenergy}
and diagram (d) of Fig.~\ref{fig:null}
need not be calculated.
As a check of our calculation we have verified
Eq.~\ref{eq:slavnovtaylor2} explicitly, including the full MSSM 
contributions.

Explicit results for $\Sigma_{W^+H^+}(p^2)$
are given in the appendix.  
The real part of $\Sigma_{W^+H^+}(m_{H^{\pm}}^2)$
fixes the $\gamma W^+H^-$ and $Z W^+H^-$ counterterms.  
The imaginary part of $\Sigma_{W^+H^+}(m_{H^{\pm}}^2)$ remains in
diagrams 1 and 2 of Fig.~\ref{fig:WHGH},
since it is not canceled by the $W^+H^+$ mixing counterterm.  
This imaginary part also determines diagram 3
of Fig.~\ref{fig:WHGH} via Eq.~\ref{eq:slavnovtaylor2}.

We have checked explicitly that all the divergences in the 1PI diagrams
that contribute to $e^+e^- \to W^+ H^-$ are canceled by the counterterms.

\section{\label{sec:numerics}Numerical results}

In this section we examine the behavior of the cross section 
for $e^+e^- \to W^+H^-$ for various choices of SUSY parameters
and evaluate the regions of parameter space in which the cross
section is large enough to be observed for $m_{H^{\pm}} > \sqrt{s}/2$.
We assume data samples of 500 fb$^{-1}$ at $\sqrt{s} = 500$ GeV
and 1000 fb$^{-1}$ at $\sqrt{s} = 1000$ GeV.  
We choose an optimistic standard of detectability to be
ten $H^{\pm}$ production events in the LC data sample.
Adding together the cross sections for $W^+H^-$ and $W^-H^+$ production,
our standard corresponds to a cross section for $W^+H^-$ production
of 0.01 fb at $\sqrt{s} = 500$ GeV and 0.005 fb at $\sqrt{s} = 1000$ GeV.
We assume that the $e^+$ beams are unpolarized.
We compare the cross sections obtained with unpolarized $e^-$ beams 
with those obtained with 80\% left- or right-polarized $e^-$ beams.
We do not make any attempt to consider backgrounds or apply cuts; this
is beyond the scope of our present analysis.

Unlike the case of the non-supersymmetric 2HDM, 
in which the top/bottom quark loops give by far the 
largest contribution to the cross section,
in the full MSSM the fermionic loops
involving charginos/neutralinos and the bosonic loops involving 
stops/sbottoms also give contributions of similar size.  
Although the stop/sbottom loops are enhanced by 
the large $H^{-}\tilde{t}_R\tilde{b}_L^*$ coupling
(which is proportional to the top quark Yukawa coupling), 
these diagrams are suppressed by higher powers of the superparticle 
masses than the fermionic loops.

Diagram 1 in Fig.~\ref{fig:susydiagrams}
and diagrams 1 and 3 in Fig.~\ref{fig:susyselfenergy} decouple
in the limit of
heavy gauginos and Higgsinos, while diagrams 3-10 in 
Fig.~\ref{fig:susydiagrams} decouple in the limit of either heavy sleptons 
or of heavy gauginos/Higgsinos.   
Squarks and sleptons contribute to diagram 
2 in Fig.~\ref{fig:susydiagrams} and 2 and 4 in 
Fig.~\ref{fig:susyselfenergy}, where stops/sbottoms give the 
largest contribution because of the large top quark Yukawa coupling.

In our numerical analysis, we use a common sfermion mass scale 
$M_{\rm SUSY}= 200$ GeV for all the squarks and sleptons.
$M_{\rm SUSY}$ is the soft SUSY-breaking mass parameter that enters 
the diagonal elements of the squark and slepton mass matrices. 
We also include the additional $D$-term contributions to squark and 
slepton masses, which is crucial for the Slavnov-Taylor identity 
(Eq.~\ref{eq:slavnovtaylor2}) to hold. 
We study two different choices for the trilinear $A$ couplings:
(I) $A_t=A_b=0$ and (II) $A_t=A_b=200$ GeV.
Along with the $\mu$ parameter and $\tan\beta$, $A_t$ determines
the left-right mixing in the stop sector, which plays an important
role for relatively light $M_{\rm SUSY}$.
In addition, the $\mu$ parameter determines the Higgsino masses. 
In most of our analysis we choose $\mu=500$ GeV;
we also consider $\mu = 100$ GeV when examining the $\tan\beta$ dependence
of the cross section.
We fix the U(1) and SU(2) gaugino mass parameters to be 
$2M_1=M_2=200$ GeV. 
Working consistently at the one-loop level, we use the tree-level
relations for the Higgs masses and mixing angles in terms of $m_{A^0}$
and $\tan\beta$.
We have verified numerically that using the radiatively corrected values for
the CP-even Higgs masses and mixing angle does not change our numerical 
results in any significant way.

In what follows, we have taken all the SUSY breaking masses for the 
squarks and sleptons to be the same for simplicity.  With 
$M_{\rm SUSY}$ as low as 200 GeV, the radiatively corrected mass
of the lightest CP-even MSSM Higgs boson $h^0$ 
lies below the current experimental Higgs search bound.  
However, it is of course possible to choose 
$M_{\rm SUSY}$ for the stop/sbottom sector to be large and/or to 
impose large left-right mixing in the stop sector so that the mass
of $h^0$ is increased above the experimental bound, while still
keeping the sleptons and first two generations of squarks relatively
light so that they give large contributions to the 
$e^+e^- \to W^{\pm}H^{\mp}$ cross section.  
In this case, the numerical results presented below will of course
change slightly, but the general conclusions from our analysis will
remain true.  A more detailed analysis of the constraints due to
the $h^0$ mass bound will be presented elsewhere~\cite{neH}.

\begin{figure*}
\resizebox{17cm}{!}{
\includegraphics{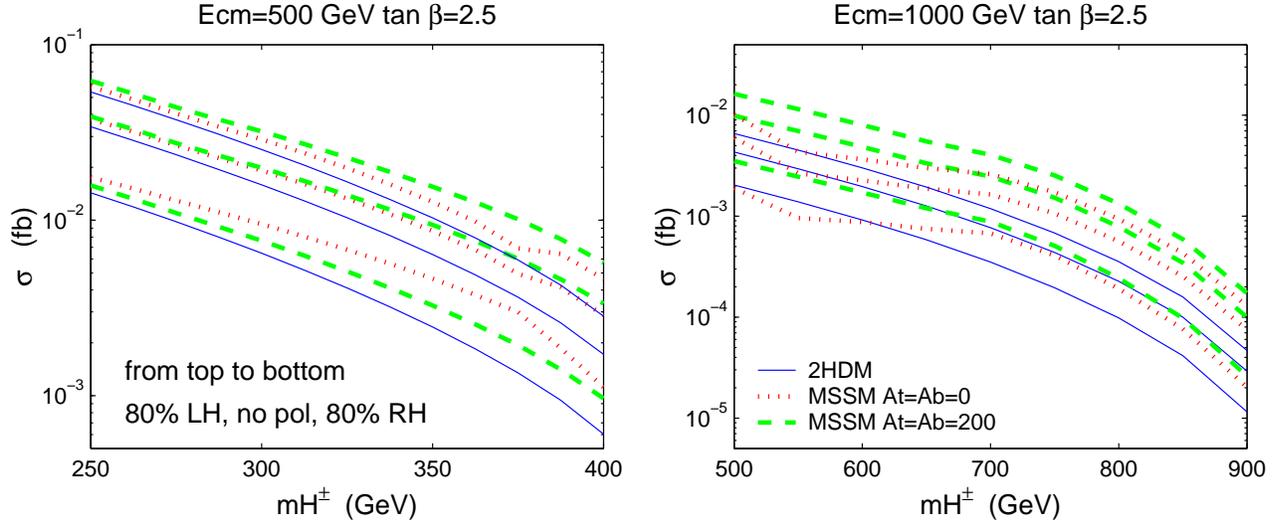}}
\caption{The $e^+e^-\rightarrow{W^+H^-}$ cross section as a function
of $m_{H^{\pm}}$ for $\tan\beta=2.5$, at 
$\sqrt{s} = 500$ GeV (left) and 1000 GeV (right).
The trilinear couplings are chosen as $A_t=A_b=0$ (dotted lines)
and 200 GeV (dashed lines).
The rest of the SUSY parameters are chosen to be
$M_{\rm SUSY}=200$ GeV, $2M_1=M_2=200$ GeV, and $\mu=500$ GeV.
The solid lines show the cross section in the non-SUSY 2HDM 
(with MSSM relations for the Higgs sector).
In each plot, the lines from top to bottom are the cross sections for 
80\% left-polarized, unpolarized, and 80\% right-polarized electrons.}
\label{fig:mHpm2.5}
\end{figure*}

\begin{figure*}
\resizebox{17cm}{!}{
\includegraphics{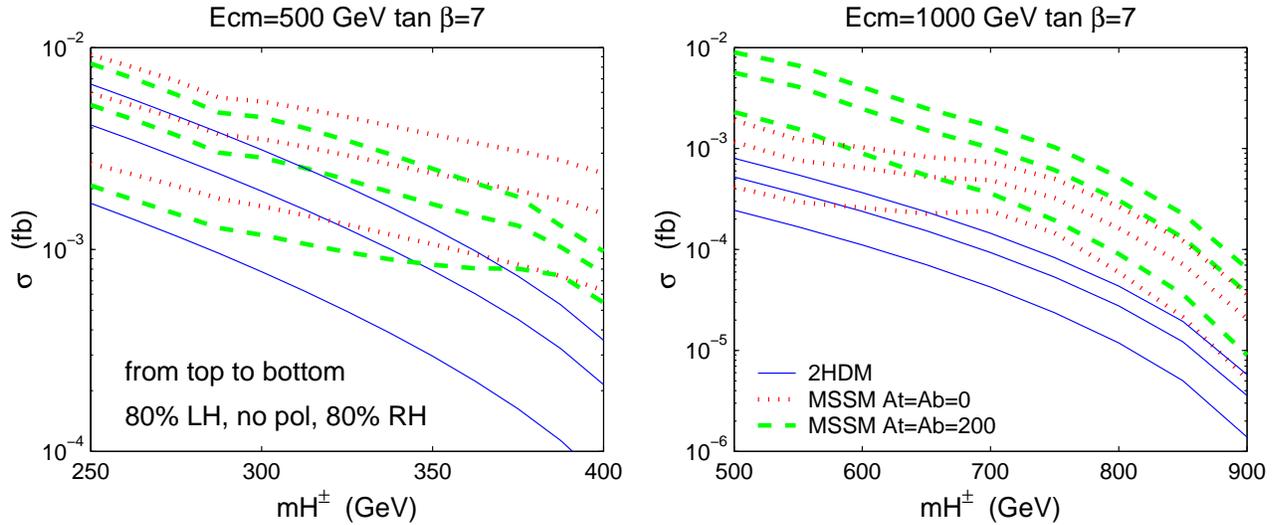}}
\caption{The $e^+e^-\rightarrow{W^+H^-}$ cross section as a function
of $m_{H^{\pm}}$ for $\tan\beta=7$.
All other input parameters and notation are the same as in 
Fig.~\ref{fig:mHpm2.5}.}
\label{fig:mHpm7}
\end{figure*}

Figures~\ref{fig:mHpm2.5} and \ref{fig:mHpm7} show the dependence of the 
$e^+e^-\rightarrow{W^+H^-}$ cross section on the charged Higgs mass
for $\tan\beta = 2.5$ and 7, respectively.
In both cases we plot cross sections for 
$\sqrt{s}=500$ GeV (left) and $\sqrt{s}=1000$ GeV (right).
Solid lines are the contributions from the non-SUSY
2HDM with the Higgs sector constrained by the MSSM mass and coupling 
relations.\footnote{The pure 2HDM contributions to 
$e^+e^-\rightarrow{W^+H^-}$ 
have been studied in Refs.~\cite{eeWH2HDM,KanemuraeeHW}.  
Our numerical results for the 2HDM are in close agreement with 
those of Refs.~\cite{eeWH2HDM,KanemuraeeHW},
after taking into account a factor of 1/4 from the average over the
initial $e^+e^-$ polarizations that was omitted in Ref.~\cite{eeWH2HDM},
and noting that the results in 
Ref.~\cite{KanemuraeeHW} are the average over the spin states
$e^+_Le^-_R$ and $e^+_Re^-_L$; for unpolarized beams the cross
sections in Ref.~\cite{KanemuraeeHW} should be divided by two.}
The dotted (dashed) lines show the cross sections in the full MSSM,
including the contributions from all the superparticles,
for $A_t=A_b=0$ ($A_t=A_b=200$ GeV).
We compare the cross sections with
80\% left-handed $e^-$ polarization, no polarization, and 80\% right-handed
$e^-$ polarization, which are denoted in each plot by the same type of lines,  
from top to bottom.  
Left-handed $e^-$ polarization always gives a larger cross section.   
The additional SUSY contributions generally enhance the cross section.  
In certain cases, for example $\sqrt{s}$=1000 GeV and
$\tan\beta=7$ (Fig.~\ref{fig:mHpm7}), the SUSY contributions increase 
the cross section by almost an order of magnitude.
The cross sections decline as $m_{H^{\pm}}$ increases; however,
reasonable cross sections can be obtained for $m_{H^{\pm}} > \sqrt{s}/2$,
especially for small $\tan\beta$.

\begin{table*}
\caption{Ten-event discovery reach in $m_{H^{\pm}}$ (in GeV) at a
LC with an 80\% left-polarized electron beam,
for $\sqrt{s}=500$ GeV (${\cal L}=500\ {\rm fb^{-1}}$) 
and 1000 GeV (${\cal L}=1000\ {\rm fb^{-1}}$). 
The corresponding reach with an unpolarized electron beam
is given in parentheses.  The SUSY parameters are as specified
in the caption of Fig.~\ref{fig:mHpm2.5}.} 
\label{tab:mHpm}
\begin{tabular}{r||c|c||c|c}\hline
&\multicolumn{2}{|c||}{$\tan\beta=2.5$}&\multicolumn{2}{|c}{$\tan\beta=7$}
\\ \hline
$\sqrt{s}$&500 GeV& 1000 GeV&500 GeV& 1000 GeV \\ \hline
2HDM &352 (327)&535 ($<$500)&$<$ 250&$<$ 500 \\
MSSM, $A_t=A_b=0$ &362 (344)&540 (512)&$<$ 250&$<$ 500 \\
MSSM, $A_t=A_b=200$ GeV&375 (347)&667 (600)&$<$ 250&581 (517) \\\hline
\end{tabular}
\end{table*}

\begin{figure*}
\resizebox{17cm}{!}{
\includegraphics{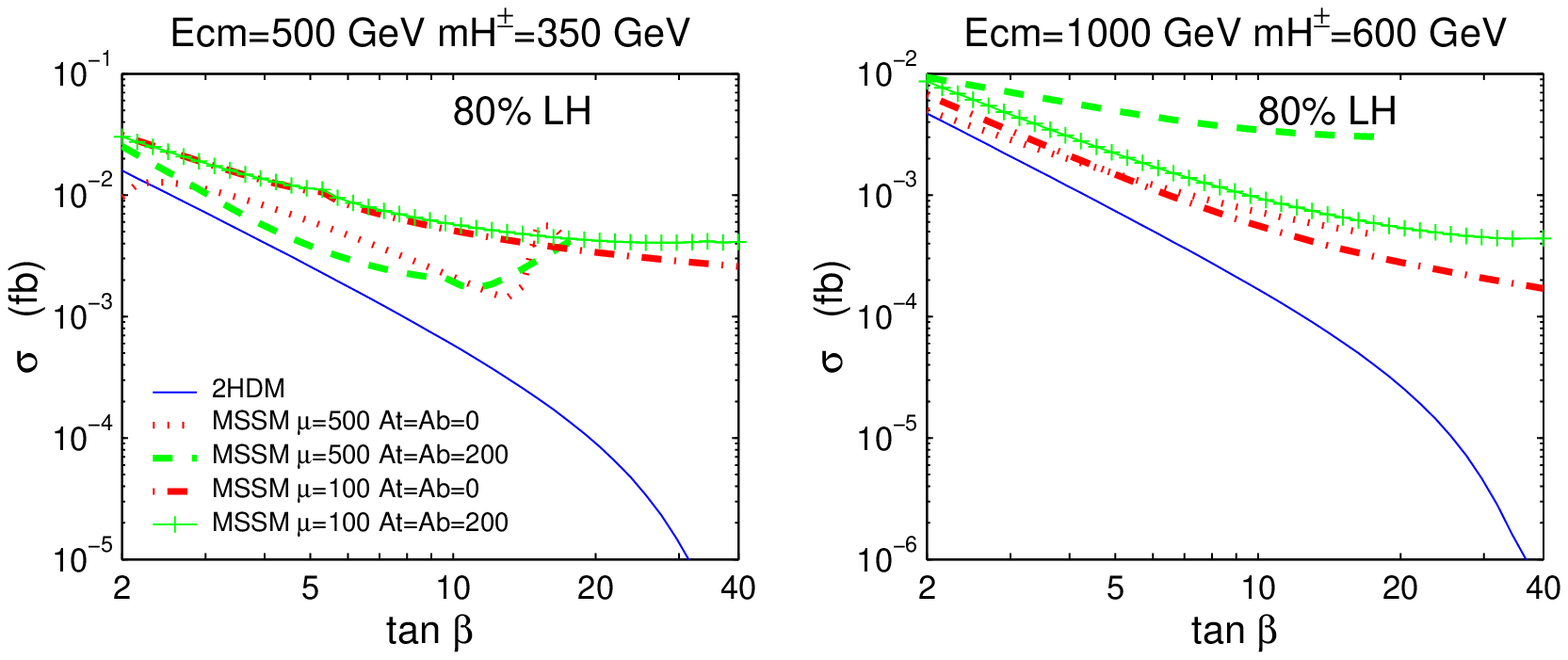}}
\caption{The $e^+e^-\rightarrow{W^+H^-}$ cross section as a function of
$\tan\beta$ for 80\% left-polarized electrons, with
$\sqrt{s} = 500$ GeV, $m_{H^{\pm}}=350$ GeV (left)
and $\sqrt{s}= 1000$ GeV, $m_{H^{\pm}}=600$ GeV (right).
Four different choices for $\mu$ and $A_t=A_b$ are shown:
$\mu=500$ GeV, $A_t=A_b=0$ (dotted line);
$\mu=500$ GeV, $A_t=A_b=200$ GeV (dashed line);
$\mu=100$ GeV, $A_t=A_b=0$ (dash-dotted line);
and $\mu=100$ GeV, $A_t=A_b=200$ GeV (``+'' line).  
The remaining SUSY parameters are
$M_{\rm SUSY}=200$ GeV and $2M_1=M_2=200$ GeV.
The solid lines show the cross section in the non-SUSY 2HDM
(with MSSM relations for the Higgs sector).}
\label{fig:beta}
\end{figure*}

The ten-event discovery reach in $m_{H^{\pm}}$ is shown
in Table~\ref{tab:mHpm} for
a 500 GeV LC (requiring a cross section of 0.01 fb)
and a 1000 GeV LC (requiring a cross section of 0.005 fb).
Using an 80\% left-polarized electron beam
generally increases the reach by at least 20 GeV compared 
to the unpolarized case.
The increase in the reach in the MSSM compared to the 2HDM is larger
at larger $\tan\beta$ or higher LC center-of-mass energy.
In the following discussion, we will consider the results for
an 80\% left-polarized electron beam.
For $\tan\beta=2.5$, a charged Higgs with mass up to 375 GeV (667 GeV) 
could be detected at a 500 GeV (1000 GeV) LC. 
For favorable SUSY parameters, the SUSY contributions
can increase the reach by about 20 GeV at $\sqrt{s} = 500$ GeV
and by more than 100 GeV at $\sqrt{s} = 1000$ GeV, compared
to the non-SUSY 2HDM.
At $\tan\beta=7$, the cross section is generally too small
to be observed at the ten-event level; however,
a reach in $m_{H^{\pm}}$ up to about 580 GeV 
is still possible at a 1000 GeV machine.  
The dependence of the cross section on $A_t$ for low 
$M_{\rm SUSY} = 200$ GeV
is due to the effects of left-right mixing in the stop sector.
For larger values of $M_{\rm SUSY}$, the squark contributions
become less important and the dependence on $A_t$ becomes
much less significant.
The cross section is also sensitive to the value of the $\mu$
parameter.  For example, lowering $\mu$ to 100 GeV increases
the ten-event discovery reach in $m_{H^{\pm}}$ to 390 GeV at $\tan\beta = 2.5$
or 300 GeV for $\tan\beta = 7$ at a 500 GeV LC, almost independent
of $A_t$ (see Fig.~\ref{fig:beta}).

Figure~\ref{fig:beta} shows the $\tan\beta$ dependence 
of the $e^+e^-\rightarrow{W^+H^-}$ cross section
for 80\% left-polarized electrons.  
We plot two different values for $\mu$, 100 and 500 GeV, to show
the $\mu$ dependence as well as the dependence on $A_t$
for small $M_{\rm SUSY} = 200$ GeV.
For $\mu=500$ GeV, values of $\tan\beta$ above 20 lead to a negative
mass-squared for the lighter sbottom and are forbidden.
From Fig.~\ref{fig:beta} we can see that while the cross section
in the 2HDM (solid line) falls rapidly with increasing $\tan\beta$,
the cross section in the MSSM experiences a much milder drop as
$\tan\beta$ increases.  
Thus the reach in $\tan\beta$ in the MSSM is larger than in the 2HDM.
At a 500 GeV machine, for $m_{H^{\pm}}=350$ GeV,
the ten-event discovery reach is $\tan\beta \leq 5.8$ in the MSSM
with favorable parameters as shown in Fig.~\ref{fig:beta},
compared to $\tan\beta < 2.5$ in the 2HDM.  
At a 1000 GeV machine, for $m_{H^{\pm}}=600$ GeV,
the ten-event discovery reach is $\tan\beta \lsim 5$ in the MSSM,
while in the 2HDM even $\tan\beta \sim 2$ is not detectable.
In Table~\ref{tab:beta} we show the maximum ten-event discovery reach in
$\tan\beta$ when $m_{H^{\pm}}$ is just above $\sqrt{s}/2$.
For a 500 GeV machine, we find that a reach up to $\tan\beta \lsim 9$
is possible for $m_{H^{\pm}} \simeq 250$ GeV.  
For a 1000 GeV machine, while in most cases we find a reach 
up to $\tan\beta \lsim 5$ for $m_{H^{\pm}} \simeq 500$ GeV, 
in certain cases a wide range of 
$\tan\beta$ values can be explored.

\begin{table*}
\caption{Ten-event discovery reach in $\tan\beta$ 
for $m_{H^{\pm}} \simeq \sqrt{s}/2$
at a LC with an 80\% left-polarized electron beam, for
$\sqrt{s}=500$ GeV (${\cal L}=500\ {\rm fb^{-1}}$)
and 1000 GeV (${\cal L}=1000\ {\rm fb^{-1}}$).
The corresponding reach with an unpolarized electron beam is given
in parentheses.  The SUSY parameters are as specified in the 
caption of Fig.~\ref{fig:beta}.}
\label{tab:beta}
\begin{tabular}{r||c|c||c|c}\hline
$\sqrt{s}$
&\multicolumn{2}{|c||}{500 GeV}&\multicolumn{2}{|c}{1000 GeV}
\\ \hline
$\mu$&100 GeV& 500 GeV&100 GeV& 500 GeV \\ \hline
2HDM &\multicolumn{2}{|c||}{5.8 (4.6)}&\multicolumn{2}{|c}{2.9 (2.3)}
\\ \cline{2-5}
MSSM, $A_t=A_b=0$ &8.6 (5.6)&6.7 (5.4)&3.8 (2.9)&3.5 (2.8)\\
MSSM, $A_t=A_b=200$ GeV&8.7 (5.9)&6.4 (4.9)&4.7 (3.4)&$\sim$ 20 \\\hline
\end{tabular}
\end{table*}

For the relatively low values of $M_{\rm SUSY}$, $M_1$ and $M_2$
used in our analysis, the dependence on $\mu$ is complicated.
We find an enhancement of the cross section over the 2HDM for most 
values of $\mu$.  
However, there are values of $\mu$ that lead to a large suppression 
in the cross section.  For example,  
at $\sqrt{s} = 500$ GeV with a left-polarized electron beam, 
$m_{H^{\pm}}=$ 350 GeV, $\tan\beta=$2.5, 
$A_t=A_b=$ 200 GeV, and $\mu$ around 800 GeV,  
there is a large
cancellation between the 2HDM and SUSY matrix elements, 
leading to a suppression of the cross section.
Furthermore, changing the relative signs of $M_1$, $M_2$ and $\mu$ can 
alter the SUSY contribution to the cross section.

\section{\label{sec:comparison}Comparison with other channels}

Single heavy Higgs boson production has been studied before
via a number of different processes, which we summarize here.
Because detailed experimental studies of almost all of these processes
are unavailable, we again choose an optimistic standard of detectability to be
10 heavy Higgs boson production events in the LC data sample.
We assume data samples of 500 fb$^{-1}$ at $\sqrt{s} = 500$ GeV
and 1000 fb$^{-1}$ at $\sqrt{s} = 1000$ GeV.  
For neutral Higgs boson production, this 10-event standard corresponds
to a cross section of 0.02 fb at $\sqrt{s} = 500$ GeV
(0.01 fb at $\sqrt{s} = 1000$ GeV).
For charged Higgs boson production, we add together the cross sections
for $H^+$ and $H^-$ production before applying the 10-event standard.
In what follows we assume that the $e^+$ and $e^-$ beams are unpolarized,
and adapt the cross sections presented in the literature accordingly.
We consider only $\tan\beta$ values above the LEP lower bound of 
2.4 \cite{LEP2} and heavy Higgs masses above $\sqrt{s}/2$.

\subsection{$e^+e^-$ collisions}

\begin{figure}
\resizebox{8.5cm}{!}{\rotatebox{270}{
\includegraphics{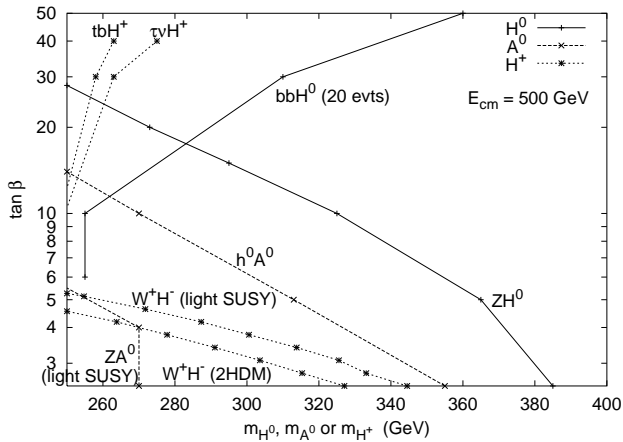}}}
\caption{Ten-event contours for single heavy Higgs boson production in
unpolarized 500 GeV $e^+e^-$ collisions, 
for 500 fb$^{-1}$ of integrated luminosity.
Solid lines show $H^0$ production via $ZH^0$ \cite{FDHiggsprod}
and $b \bar b H^0$ \cite{bbHAHpm} (the $b \bar b H^0$ line
is a 20-event contour; see text).
Long-dashed lines show $A^0$ production via 
$h^0A^0$ \cite{FDHiggsprod} 
and $ZA^0$ \cite{eeZAMSSM}.
Short-dashed lines show $H^+$ production via
$\bar t b H^+$ and $\tau \bar \nu H^+$ \cite{KanemuraReview};
also shown are our results for $W^+H^-$ production in the 2HDM (lower curve)
and full MSSM with light superpartners (upper curve).
On the $x$-axis we plot the mass of the relevant heavy Higgs boson.
See text for details.
}
\label{fig:shhp500}
\end{figure}

\begin{figure}
\resizebox{8.5cm}{!}{\rotatebox{270}{
\includegraphics{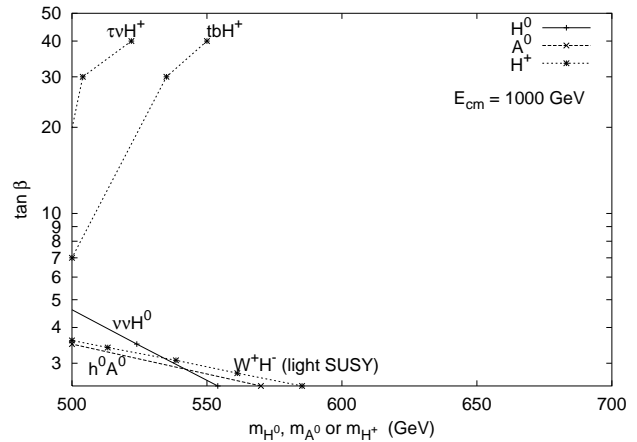}}}
\caption{As in Fig.~\ref{fig:shhp500} but for 
1000 GeV $e^+e^-$ collisions and 1000 fb$^{-1}$ of integrated luminosity.
The solid line shows $\nu \bar \nu H^0$ production.
The long-dashed line shows $h^0A^0$ production \cite{FDHiggsprod}.
Short-dashed lines show $H^+$ production via
$\bar t b H^+$ and $\tau \bar \nu H^+$ \cite{KanemuraReview};
also shown are our results for $W^+H^-$ production in the 
full MSSM with light superpartners.
}
\label{fig:shhp1000}
\end{figure}

The 10-event reach for various single heavy Higgs boson production 
modes in 500 GeV and 1000 GeV $e^+e^-$ collisions is shown in 
Figs.~\ref{fig:shhp500} and \ref{fig:shhp1000}, respectively.

The heavy CP-even neutral
Higgs boson $H^0$ can be produced via
Higgsstrahlung, $e^+e^- \to ZH^0$; $W$-boson fusion,
$e^+e^- \to \nu \bar \nu H^0$; and $Z$-boson fusion,
$e^+e^- \to e^+ e^- H^0$.  The cross sections for these
processes are suppressed relative to the 
corresponding SM Higgs production cross
sections by $\cos^2(\beta - \alpha)$, where $\alpha$ is the mixing angle
that diagonalizes the CP-even neutral Higgs boson mass-squared matrix.
In the decoupling limit \cite{decoupling}, 
$\cos^2(\beta - \alpha) \propto m_Z^4/m_{A^0}^4$,
so these cross sections decrease rapidly as $m_{A^0}$ increases.  
At a 500 GeV LC, $Z H^0$ production yields $\geq 10$ events
well beyond $m_{H^0} \simeq 0.5 \sqrt{s}$, out to $m_{H^0} \lsim 380$ GeV 
and for low to moderate $\tan\beta \lsim 25$ (see Fig.~\ref{fig:shhp500}). 
The contour shown in Fig.~\ref{fig:shhp500} was produced
using the program FeynHiggsXS~\cite{FDHiggsprod}, which includes the full 
Feynman-diagrammatic corrections to this process at one-loop
as well as two-loop corrections to the Higgs masses and 
mixing.\footnote{The FeynHiggsXS~\cite{FDHiggsprod} input parameters
were chosen as $M_{\rm SUSY} = 1000$ GeV for all three generations, 
$M_2 = 200$ GeV, $M_{\tilde g} = 500$ GeV, $\mu = -200$ GeV, 
and $A_t = A_b = 2 M_{\rm SUSY}$ to yield the maximal mixing scenario.  
$M_1$ was fixed by the GUT relation to $M_2$.}
At a 1000 GeV LC, the cross section for $Z H^0$ production is 
too small to be of use. 
The cross section for Higgs boson production via $W$-boson fusion
typically becomes more important as $\sqrt{s}$ increases
(see, {\it e.g.}, Ref.~\cite{Orangebook}).
While the cross section for $\nu \bar \nu H^0$ at $\sqrt{s} = 500$ GeV
is too small to be of use, at $\sqrt{s} = 1000$ GeV it yields
$\geq 10$ events for $m_{H^0} \lsim 550$ GeV and $\tan\beta \lsim 4.5$
(see Fig.~\ref{fig:shhp1000}).\footnote{These results
for $e^+e^- \to \nu \bar \nu H^0$ are based on the $\alpha_{\rm eff}$ 
approximation from the program subhpoldm \cite{subhpoldm},
with the same MSSM input parameters as used in the FeynHiggsXS computation
above.  A full Feynman-diagrammatic calculation of 
the radiative corrections to this process is not yet available.}
Finally, the $Z$-boson fusion cross section is always about a factor
of ten smaller than the $W$-boson fusion cross section, 
so we do not expect it to be useful here.

In addition to the kinematically limited production mode
$e^+e^- \to H^0 A^0$, the heavy CP-odd neutral Higgs boson can
be produced in association with the light $h^0$, via $e^+e^- \to h^0 A^0$. 
As above, the cross section for this production mode is suppressed
by $\cos^2(\beta - \alpha)$.  Using FeynHiggsXS~\cite{FDHiggsprod}
with input parameters as given above, we find that 
at a 500 GeV LC $h^0A^0$ production yields $\geq 10$ events
for $m_{A^0} \lsim 350$ GeV and low to moderate $\tan\beta \lsim 14$
(see Fig.~\ref{fig:shhp500}); at 1000 GeV the 10-event reach is 
$m_{A^0} \lsim 560$ GeV and $\tan\beta \lsim 3.5$ 
(see Fig.~\ref{fig:shhp1000}).

The heavy MSSM Higgs bosons can also be produced in association with
pairs of third-generation fermions: $e^+e^- \to b \bar b H^0$, $b \bar b A^0$,
$\tau^- \bar \nu H^+$, and $\bar t b H^+$.  The cross sections for the first
three of these processes are strongly enhanced at large $\tan\beta$, and 
the fourth is enhanced at both large and small 
$\tan\beta$.\footnote{The first two processes are useful for measuring 
$\tan\beta$ at large $\tan\beta$ and low $m_{A^0}$~\cite{bbAtanbeta}.}
At a 500 GeV LC, the process $e^+e^- \to b \bar b H^0$ is especially
promising: it yields $\gsim 20$ events for $m_{H^0} \lsim 360$ GeV
and large $\tan\beta \gsim 10$ \cite{bbHAHpm,bbHA}; 
this 20 event contour is shown in 
Fig.~\ref{fig:shhp500}.\footnote{Results were given in 
Refs.~\cite{bbHAHpm,bbHA} for tree-level 
$e^+e^- \to b\bar b H^0$, $b \bar b A^0$ cross sections 
down to 0.1 fb, which yields 50 signal events, using $m_b = 4.25$ GeV.
We take into account the dominant QCD corrections 
\cite{bbHQCD} by scaling these
cross sections by $(\overline{m}_b(m_{H^+})/m_b^{\rm pole})^2 \simeq 0.4$
for $m_{H^+} \sim 250-350$ GeV, so that the cross section of 0.1 fb
in Refs.~\cite{bbHAHpm,bbHA} corresponds to roughly 20 signal events
after QCD corrections are included.}
The process $e^+e^- \to b \bar b A^0$ yields less than 20 events
for $m_{A^0} > 250$ GeV~\cite{bbHAHpm,bbHA}.
At a 1000 GeV LC, neither $b \bar b H^0$ nor $b \bar b A^0$ production
yields more than 10 events for $m_{H^0,A^0} > 500$ GeV~\cite{bbHAHpm}.

At a 500 GeV LC, both $\tau^- \bar \nu H^+$ and 
$\bar t b H^+$ production yield 
$\geq 10$ events at large $\tan\beta \sim 40$ for $m_{H^{\pm}} \lsim 270$ GeV,
with $\tau^- \bar \nu H^+$ production having a slightly larger cross
section \cite{KanemuraReview,bbHAHpm,bbHpm} (see Fig.~\ref{fig:shhp500}).
At a 1000 GeV LC, $\bar t b H^+$ production is more promising, due to
the larger phase space available; it yields $\geq 10$ events at large
$\tan\beta \sim 40$ for $m_{H^{\pm}} \lsim 550$ GeV, while 
$\tau^- \bar \nu H^+$ production gives a reach of only $m_{H^{\pm}} \lsim 520$
GeV \cite{KanemuraReview,bbHAHpm} 
(see Fig.~\ref{fig:shhp1000}).\footnote{The QCD corrections to
$e^+e^- \to \bar t b H^+$ were recently computed in Ref.~\cite{tbHQCD} 
and were found to reduce the cross section at large $\tan\beta$; 
in particular, for $\sqrt{s} = 500$ GeV and $\tan\beta=40$, the QCD
corrections reduce the 10-event reach by about 8 GeV \cite{tbHQCD}.}

Finally we consider single heavy MSSM Higgs production modes that are zero
at the tree level but arise at one loop.
The process $e^+e^- \to Z A^0$ has been calculated in both the 
2HDM \cite{eeZA2HDM} and in the full 
MSSM \cite{eeZAMSSM}.\footnote{A significant contribution to 
$e^+e^- \to Z A^0$ comes from the loop-induced $A^0 ZZ$ vertex, 
computed in Ref.~\cite{AZZvertex}.}
In the 2HDM the cross section is too small to be of 
use at either $\sqrt{s} = 500$ or 1000 GeV \cite{eeZA2HDM}. 
In the full MSSM, light charginos and sleptons with masses of order
200 GeV can enhance the cross section by an order of magnitude or more;
this leads to $\geq 10$ events at a 500 GeV LC
for $m_{A^0} \lsim$ 270 GeV and $\tan\beta \lsim 5$ 
\cite{eeZAMSSM} (see Fig.~\ref{fig:shhp500}).
This enhancement disappears once the chargino and slepton masses exceed about
500 GeV.  For $\sqrt{s} = 1000$ GeV, even with light SUSY particles
the cross section is too small to be of interest here.
The similar process $e^+e^- \to \gamma A^0$ was considered in
Refs.~\cite{gammaA,eeZA2HDM,FGL}; however, even including
the enhancement of the cross section from the contributions of light SUSY 
particles, this process yields too few events to be of use at 
either $\sqrt{s} = 500$ or 1000 GeV.

Comparing the process $e^+e^- \to W^+H^-$ to the various processes 
described above, we see that $W^+H^-$ production is the only channel
in $e^+e^-$ collisions
analyzed to date that yields $\geq 10$ events containing charged Higgs bosons
at low $\tan\beta$ values for $m_{H^{\pm}}\geq \sqrt{s}/2$.  
In the 2HDM with MSSM relations imposed upon the Higgs sector, $W^+H^-$ 
yields $\geq 10$ events for $m_{H^{\pm}} \lsim 325$ GeV and 
$\tan\beta \lsim 4.5$ at $\sqrt{s} = 500$ GeV (Fig.~\ref{fig:shhp500}), 
while for $\sqrt{s} = 1000$ GeV the cross section in the 2HDM is too small
to be of use.
If the MSSM constraint on the Higgs sector is relaxed,
then the $e^+e^- \to W^+H^-$ cross section can be enhanced in some regions of
2HDM parameter space due to large triple-Higgs couplings \cite{KanemuraeeHW}.
Under the requirement that the Higgs self-couplings remain perturbative,
this enhancement can increase the reach
at $\sqrt{s} = 500$ GeV up to $\tan\beta \lsim 6.5$
for $m_{H^{\pm}} \lsim 280$ GeV \cite{KanemuraeeHW,eeWH2HDM}.
Including the full MSSM contributions,\footnote{The MSSM parameters
were chosen here as $M_{\rm SUSY} = 1000$ GeV for the third generation
squarks, $M_{\rm SUSY} = 200$ GeV for the rest of the squarks and sleptons,
$M_2 = 2 M_1 = 200$ GeV, $A_t = A_b = 2000$ GeV ({\it i.e.}, the maximal
mixing scenario), and $\mu = 200$ GeV for $\sqrt{s} = 500$ GeV and
$\mu = 500$ GeV for $\sqrt{s} = 1000$ GeV.  The tree-level relations
were used to determine the MSSM Higgs masses and couplings; we checked
numerically that using the radiatively-corrected MSSM Higgs masses and
couplings does not change our results significantly \cite{neH}.}
the reach increases to $m_{H^{\pm}} \lsim 345$ GeV and $\tan\beta \lsim 5$ 
at $\sqrt{s} = 500$ GeV (Fig.~\ref{fig:shhp500}), 
and $m_{H^{\pm}} \lsim 580$ GeV and 
$\tan\beta \lsim 3.5$ at $\sqrt{s} = 1000$ GeV (the latter reach is 
comparable to that of $h^0A^0$ production; see Fig.~\ref{fig:shhp1000}).

\subsection{$\gamma\gamma$ and $e^- \gamma$ collisions}

If the $e^+e^-$ LC is converted into a photon collider through Compton 
backscattering of intense laser beams, 
the neutral heavy Higgs bosons $H^0$ and $A^0$ can be singly produced in the 
$s$-channel through their loop-induced couplings to photon pairs.  
This process appears to be very promising for detecting 
$H^0$ and $A^0$ with masses above $\sqrt{s}/2$ 
and moderate $\tan\beta$ values between 2.5 and 10
\cite{gamgamAsner,gamgamMuhlleitner,gamgamGunionHaber}.
In particular, a recent realistic simulation of signal and 
backgrounds \cite{gamgamAsner} showed that
a 630 GeV $e^+e^-$ LC running in $\gamma\gamma$ mode
for three years would allow $H^0$, $A^0$ detection 
over a large fraction of the LHC wedge region (in which the heavy MSSM
Higgs bosons would not be discovered at the LHC)
for $m_{A^0}$ up to the photon-photon energy limit of 
$\sim 500$ GeV.  
At a 1000 GeV LC, the mass reach is likely to be above 
600 GeV \cite{gamgamMuhlleitner}.

The cross sections for production of $\tau^- \bar \nu H^+$ and 
$\bar t b H^+$ in $\gamma\gamma$ collisions \cite{CPYuan} are expected to be 
larger than the corresponding cross sections in $e^+e^-$ collisions
at large $\tan\beta$.
In particular, if 1000 fb$^{-1}$ of $e^+e^-$ luminosity at 
$\sqrt{s_{ee}} = 1000$ GeV is devoted
to $\gamma\gamma$ collisions, $\tau^- \bar \nu H^+$ production
yields $\geq 10$ events for $m_{H^{\pm}}$ up to almost 700 GeV
at large $\tan\beta \simeq 30$ \cite{CPYuan}.

Production of $W^+H^-$ in $\gamma\gamma$ collisions also 
occurs at the one-loop level.  The cross section for this process has 
been computed in the non-supersymmetric 2HDM and the full
MSSM in Ref.~\cite{ggHW}.
If 1000 fb$^{-1}$ of $e^+e^-$ luminosity at 1000 GeV 
is devoted to $\gamma\gamma$ collisions, $\geq 10$ events would be produced
in the 2HDM for $m_{H^{\pm}} \lsim 570$ GeV at $\tan\beta = 2$; 
the cross section falls with increasing $\tan\beta$.
In the full MSSM with light superparticles the cross section is 
enhanced, and this mode provides $\geq 10$ events for 
$m_{H^{\pm}}$ above 600 GeV at $\tan\beta = 2$ ($m_{H^{\pm}} \lsim 520$ GeV
at $\tan\beta = 6$) \cite{ggHW}.  
Assuming comparable luminosities,
$\gamma\gamma \to W^+H^-$ is competitive with $e^+e^- \to W^+H^-$ in the MSSM.
This implies that spending some of the $e^+e^-$ luminosity on running
in photon collider mode would not significantly impact the number of 
$W^+H^-$ signal events collected.

Finally, if the LC is run in $e^- \gamma$ mode, the process
$e^- \gamma \to \nu H^-$ is possible.  The cross section for this process
has been computed in the 2HDM \cite{Kanemuraegamma} and found to be  
relatively independent of $m_{H^{\pm}}$
all the way up to the kinematic threshold for 
parent $e^+e^-$ center-of-mass energies of 500 or 1000 GeV.  
Unfortunately, with the typical expected $e^- \gamma$ 
luminosity of 100 fb$^{-1}$ \cite{Teslaegamma},
the cross section for this process in the 2HDM is too small to be of interest
\cite{Kanemuraegamma}.
This process could become promising in the MSSM if its cross section
is enhanced by the contributions of light superpartners \cite{neH}, or if the
$e^- \gamma$ luminosity is increased.

\section{\label{sec:conclusions}Conclusions}

We studied the single charged Higgs production process 
$e^+e^-\rightarrow{W}^\pm{H}^\mp$ at a 500 GeV and a 1000 GeV linear collider, 
including the complete MSSM contributions at one loop.  
At small $\tan\beta$, the reach in $m_{H^{\pm}}$ from this process
extends above the kinematic threshold ($m_{H^{\pm}} \simeq \sqrt{s}/2$)
for $e^+e^- \to H^+H^-$ pair production.
We found that light SUSY particles with masses of order
200 GeV enhance the cross section in most 
of the parameter space.
At small $\tan\beta$, an increase in the reach in $m_{H^{\pm}}$ 
by about 20 GeV in the full MSSM compared
to that in the 2HDM is possible 
at a 500 GeV machine, and by more than 100 GeV at a 1000 GeV machine.
In particular, with favorable SUSY parameters and $\tan\beta = 2.5$, 
this process yields more than 10 $W^{\pm}H^{\mp}$ events
for $m_{H^{\pm}} \lsim 350$ GeV in 500 fb$^{-1}$ at a 500 GeV collider,
or $m_{H^{\pm}} \lsim 600$ GeV in 1000 fb$^{-1}$ at a 1000 GeV collider,
assuming unpolarized beams.
At large $\tan\beta$, an order of magnitude enhancement of the 
cross section compared to the 2HDM is possible,
so that the cross section suffers only a moderate decrease with increasing
$\tan\beta$.  The reach in $\tan\beta$ is therefore enhanced;
with favorable SUSY parameters and $m_{H^{\pm}} \simeq \sqrt{s}/2$,
this process yields more than 10 $W^{\pm}H^{\mp}$ events for
$\tan\beta \lsim 6$ at a 500 GeV collider.
We also found that an 80\% left-polarized electron beam enhances
the cross section by about 50\% and improves
the discovery reach by at least an additional 25 GeV;
in particular, the reach at $\tan\beta = 2.5$ improves to 
$m_{H^{\pm}} \lsim 375$ GeV at a 500 GeV collider,
or $m_{H^{\pm}} \lsim 670$ GeV at a 1000 GeV collider, and the reach in
$\tan\beta$ improves to $\sim 8.5$ at a 500 GeV collider.

The process $e^+e^- \to W^+H^-$ in the MSSM with light superpartners
is very promising compared to the other single charged Higgs boson 
production modes that have been considered to date.
At the $e^+e^-$ collider, $W^+H^-$ production at low $\tan\beta$
is complementary to $\tau^- \bar \nu H^+$ and $\bar t b H^+$
production at high $\tan\beta$.
At an $e^- \gamma$ collider, the cross section for $\nu H^-$ 
production in the 2HDM is too small to be seen 
with a typical integrated luminosity of 100 fb$^{-1}$ in the high-energy
part of the $\gamma$ spectrum; however,
this process could become promising in the MSSM if its cross section
is enhanced by the contributions of light superpartners, or if the
$e^- \gamma$ luminosity is increased.
Finally, if 1000 fb$^{-1}$ of $e^+e^-$ luminosity at 1000 GeV is devoted
to $\gamma\gamma$ collisions, the process $\gamma\gamma \to W^+H^-$
in the MSSM will be competitive to $e^+e^- \to W^+H^-$.  

A few additional processes that have not yet been computed may be 
promising for single heavy Higgs boson production at an $e^+e^-$ collider.
The behavior of the SM Higgs production cross sections leads
us to expect that weak boson fusion processes will naturally have larger
cross sections than Higgsstrahlung-type processes at $\sqrt{s} \sim 1000$ GeV.
The process $e^+e^- \to \bar \nu e^- H^+$ may thus be promising~\cite{FGLS}.
The process $e^+e^- \to \nu \bar \nu A^0$ may also be promising;
however, preliminary results for the top/bottom quark loop contributions 
in the 2HDM indicate that this process is 
too small to be observed for $m_{A^0} > \sqrt{s}/2$ \cite{FGL}.  
This process could become observable if
the addition of the full 2HDM and MSSM contributions enhances
the cross section sufficiently \cite{FGLS}.

\begin{acknowledgments}
We are grateful to Abdesslam Arhrib, David Asner, Marcela Carena, Tom Farris,
Jack Gunion, Howie Haber,
Sven Heinemeyer, Uli Nierste, Michel Capdequi-Peyran\`ere,
Tilman Plehn, Dave Rainwater and Tim Tait for helpful conversations.
We also thank the Aspen Center for Physics, where this project was begun.
HEL thanks the Physics Department at Caltech for their 
hospitality during part of this work.
Fermilab is operated by Universities Research Association Inc.\
under contract no.~DE-AC02-76CH03000 with the U.S. Department of
Energy.  S.S. is supported by the DOE under grant DE-FG03-92-ER-40701 and 
by the John A. McCone Fellowship.
\end{acknowledgments}

\appendix

\section{\label{app:conventions}Notation and conventions}
For couplings and Feynman rules we follow the conventions of
Ref.~\cite{HaberKane}.
We follow the notation of Ref.~\cite{eeWH2HDM} for the one-loop integrals.
The one-point integral is:
\begin{equation}
 	\frac{i}{16 \pi^2}A(m^2) = \int \! \frac{d^Dq}{(2\pi)^D}
	\frac{1}{(q^2 - m^2)},
\end{equation}
where $D$ is the number of dimensions.
The two-point integrals are:
\begin{eqnarray}
	&&\frac{i}{16\pi^2}\left\{B_0, k^{\mu}B_1 \right\}(k^2,m_0^2,m_1^2)
	= \\
	&&\hskip1cm \int \! \frac{d^Dq}{(2\pi)^D} 
	\frac{ \{1, q^{\mu} \} }{(q^2 - m_0^2)((q+k)^2 - m_1^2)}.
	\nonumber
\end{eqnarray}
The three-point integrals are:
\begin{eqnarray}
	&&\frac{i}{16 \pi^2}\left\{ C_0, C^{\mu}, C^{\mu\nu} \right\}
	= \\
	&&\hskip0cm \int \! \frac{d^Dq}{(2\pi)^D}
	\frac{ \{1, q^{\mu}, q^{\mu}q^{\nu} \} }
	{(q^2 - m_0^2)((q+k_1)^2 - m_1^2)((q+k_2)^2 - m_2^2)},
	\nonumber
\end{eqnarray}
where the tensor integrals are decomposed in terms of scalar components
as
\begin{eqnarray}
	C^{\mu} &=& k_1^{\mu}C_1 + k_2^{\mu}C_2  \nonumber \\
	C^{\mu\nu} &=& g^{\mu\nu} C_{00} + k_1^{\mu}k_1^{\nu}C_{11}
	+ k_2^{\mu}k_2^{\nu}C_{22} \nonumber \\
	&& + (k_1^{\mu}k_2^{\nu} + k_2^{\mu}k_1^{\nu})C_{12}.
\end{eqnarray}
The arguments of the scalar three-point integrals are 
$(k_1^2,(k_2-k_1)^2,k_2^2,m_0^2,m_1^2,m_2^2)$.
The four-point integrals are:
\begin{widetext}
\begin{eqnarray}
	&&\frac{i}{16 \pi^2}\left\{ D_0, D^{\mu}, D^{\mu\nu} \right\}
	=\hskip0cm \int \! \frac{d^Dq}{(2\pi)^D}
	\frac{ \{1, q^{\mu}, q^{\mu}q^{\nu} \} }
	{(q^2 - m_0^2)((q+k_1)^2 - m_1^2)((q+k_2)^2 - m_2^2)((q+k_3)^2-m_3^2)},
\end{eqnarray}
\end{widetext}
where the tensor integrals are decomposed in terms of scalar components
as
\begin{eqnarray}
	D^{\mu} &=& k_1^{\mu}D_1 + k_2^{\mu}D_2 + k_3^{\mu}D_3  \nonumber \\
	D^{\mu\nu} &=& g^{\mu\nu} D_{00} + k_1^{\mu}k_1^{\nu}D_{11}
	+ k_2^{\mu}k_2^{\nu}D_{22} + k_3^{\mu}k_3^{\nu}D_{33}\nonumber \\
	&&+ (k_1^{\mu}k_2^{\nu} + k_2^{\mu}k_1^{\nu})D_{12}
	+ (k_1^{\mu}k_3^{\nu} + k_3^{\mu}k_1^{\nu})D_{13}
	\nonumber \\
	&&+ (k_2^{\mu}k_3^{\nu} + k_3^{\mu}k_2^{\nu})D_{23}.
\end{eqnarray}
The arguments of the scalar four-point integrals are 
$(k_1^2, (k_2-k_1)^2, (k_3-k_2)^2, k_3^2, k_2^2, 
(k_1-k_3)^2, m_0^2, m_1^2, m_2^2, m_3^2)$.


\section{2HDM contributions}
Explicit results for the matrix elements for $e^+e^- \to W^+H^-$ 
in the 2HDM were given in Ref.~\cite{eeWH2HDM}.  We have calculated
all the 2HDM diagrams independently and give the formulae here 
for completeness.  After correcting a few typographical 
errors \cite{Arhribprivate} in the formulae of Ref.~\cite{eeWH2HDM},
we find agreement with all of their results except for 
Figs.~\ref{fig:WHGH}.2 and \ref{fig:WHGH}.3, which
differ by an overall sign.

The quark triangles that appear in Figs.~\ref{fig:2HDMdiagrams}.1 and
\ref{fig:2HDMdiagrams}.2 were also computed
in Ref.~\cite{PHI} and agree with our results.  (Note that
Ref.~\cite{PHI} uses the convention $\varepsilon_{0123}=1$, which is
opposite to our convention.)

For convenience, we list here our notation for the gauge and Yukawa
coupling coefficients used in this section.

The photon coupling coefficients to leptons/quarks are:
\begin{equation}
g_{\gamma}^{fL}=g_{\gamma}^{fR}=-e_f,
\end{equation}
where the electric charges are $e_{\nu} = 0$, $e_e = -1$, $e_u = 2/3$,
and $e_d = -1/3$.

The $Z$ boson coupling coefficients to leptons/quarks are:
\begin{equation}
g_{Z}^{fL}=(-T_3+e_f\sw^2)/s_Wc_W,\ \ \ 
g_{Z}^{fR}=(e_f\sw^2)/s_Wc_W,
\end{equation}
where $T_3=1/2$ for $\nu$, $u$ and $T_3=-1/2$ for $e$, $d$.

For the $W$ boson coupling to leptons/quarks we define: 
\begin{equation}
\gw=-1/\sqrt{2}\sw.
\end{equation}

The $H^{\pm}$ and $G^{\pm}$ coupling coefficients to top/bottom quarks are 
(the couplings to leptons and first and second generation quarks 
are small and taken to be zero):

\begin{widetext}

\begin{center}
\begin{tabular}{|c|c|c|c|} \hline
$\yl$&$\yr$&$\ygl$&$\ygr$  \\ \hline
$\mt\cot\beta/\sqrt{2}\mw\sw$ &  
$\mb\tan\beta/\sqrt{2}\mw\sw$ &  
$-\mb/\sqrt{2}\mw\sw$ &  
$\mt/\sqrt{2}\mw\sw$  \\  \hline
\end{tabular}
\end{center}

The coefficients for couplings of three Higgs bosons are:
\begin{center}
\begin{tabular}{|c|c|} \hline
$\ghag$&$-i \mw/2\sw$ \\
$\ghhh$&$-\mw\sba/\sw - \mz \cos{2}\beta\sin(\beta+\alpha)/2\sw\cw$  \\
$\ghHh$&$-\mw\cba/\sw + \mz\cos{2}\beta\cos(\beta+\alpha)/2\sw\cw$  \\
$\ghhg$&$\mw\cba/2\sw - \mz\sin2\beta\sin(\beta+\alpha)/2\sw\cw$  \\
$\ghHg$&$-\mw\sba/2\sw + \mz\sin2\beta\cos(\beta+\alpha)/2\sw\cw$  \\
$\gghg$&$\mz\cos{2}\beta\sin(\beta+\alpha)/2\sw\cw$  \\
$\ggHg$&$-\mz\cos{2}\beta\cos(\beta+\alpha)/2\sw\cw$  \\  \hline
\end{tabular}
\end{center}
where $\sba=\sin(\beta-\alpha)$ and $\cba=\cos(\beta-\alpha)$, and 
we have imposed the MSSM relations on the $H^+H^-h^0$, $H^+H^-H^0$ couplings.

We now list our results for the 2HDM diagrams.

Fig.~\ref{fig:2HDMdiagrams}.1:
\begin{eqnarray}
{\cal{M}}&=&\frac{2\alpha^2{N}_c{g}_W}{s-\mvtwo}\left\{
[-(m_t\gvdl\yl+m_b(\gvdl-\gvdr)\yr)B_0
+2\gvdl(m_t\yl+m_b\yr)C_{00}\right. \nonumber \\
&&\left.
+(-\gvdl(m_t\yl+m_b\yr)\mwtwo+(m_t\gvdl\yl-m_b\gvdr\yr)k_1\cdot{k}_2)C_1
\right.\nonumber \\
&&\left.
+(\gvdl(m_t\yl+m_b\yr)k_1\cdot{k}_2- (m_t\gvdl\yl-m_b\gvdr\yr)\mhpmtwo)C_2
\right. \nonumber \\
&&\left.
+(-m_t^3\gvdl\yl-m_t^2m_b(\gvdl-\gvdr)\yr+m_tm_b^2\gvdr\yl+
m_t\gvdl\yl{k}_1\cdot{k}_2)C_0][\gver{\cal{A}}_1+\gvel{\cal{A}}_2]
\right. \nonumber \\
&&\left. 
+[-2\gvdl(m_t\yl+m_b\yr)(C_{12}+C_{22})-\gvdl(3m_t\yl+m_b\yr)C_2
-(m_t\gvdl\yl-m_b\gvdr\yr)C_1
\right. \nonumber \\
&&\left.
-m_t\gvdl\yl{C}_0]
[\gver(\athree+\afive)+\gvel(\afour+\asix)]
\right.\nonumber \\
&&\left.+[m_t\gvdl\yl{C}_0+(m_t\gvdl\yl+m_b\gvdr\yr)C_1
+\gvdl(m_t\yl+m_b\yr)C_2]
(\gver\aseven+\gvel\aeight)\right\},
\end{eqnarray} 
with the arguments for the integral functions as 
$B(s,\mbtwo,\mbtwo)$, $C(\mwtwo,s,\mhpmtwo,\mttwo,\mbtwo,\mbtwo)$.

Fig.~\ref{fig:2HDMdiagrams}.2:
Similar to Fig.~\ref{fig:2HDMdiagrams}.1 
with the exchange of 
\begin{equation}
\mt\leftrightarrow\mb, \ \ \gvdl\leftrightarrow\gvul,\ \ \ 
\gvdr\leftrightarrow\gvur, \ \ \ \yl\leftrightarrow\yr;\ \ \ 
\aseven, \aeight\ {\rm terms\ flip\ sign}.\nonumber 
\end{equation}

Fig.~\ref{fig:2HDMdiagrams}.3:
\begin{eqnarray}
{\cal{M}}&=&
\frac{\alpha^2}{s-\mvtwo}g_{VWW}g_{WWS}g_{HWS}
\nonumber \\
&\times& \left\{ \left[-B_0-(s+\mstwo-\mwtwo)C_0-(s+\mwtwo-\mhpmtwo)C_1
+(s-\mwtwo+\mhpmtwo)C_2+C_{00}\right](\gver\aone+\gvel\atwo) \right.
\nonumber \\
&& \left. + (4C_1+C_2-C_{12}-C_{22})
\left[\gver(\athree+\afive)+\gvel(\afour+\asix)\right] \right\},
\end{eqnarray}
with the arguments for the integral functions as 
$B(s,\mwtwo,\mwtwo)$, 
$C(\mwtwo,s,\mhpmtwo,\mstwo,\mwtwo,\mwtwo)$.
The couplings are given in the following table:
\begin{center}
\begin{tabular}{|c|c|c|c|c|}\hline
$S$ & $g_{\gamma{W}{W}}$ & $g_{ZWW}$ & $g_{HWS}$ & $g_{WWS}$ \\ \hline 
$h^0$ & $-1$ & $-\cw / \sw$ & $\cba / 2\sw$  & $\mw \sba / \sw$\\
$H^0$ & $-1$ & $-\cw / \sw$ & $-\sba / 2\sw$ & $\mw \cba / \sw$\\
\hline 
\end{tabular}
\end{center}

Fig.~\ref{fig:2HDMdiagrams}.4:
\begin{eqnarray}
{\cal{M}}&=&
\frac{2\alpha^2}{s-\mztwo}g_{ZS_iS_j}g_{HWS_i}g_{WWS_j}\left\{
C_{00}(\gver\aone+\gvel\atwo) \right. \nonumber \\
&-&\left.(2C_0+2C_1+3C_2+C_{12}+C_{22})
\left[\gver(\athree+\afive)+\gvel(\afour+\asix)\right]\right\},
\end{eqnarray}
with the arguments for the integral functions as 
$C(\mwtwo,s,\mhpmtwo,\mwtwo,\msjtwo,\msitwo)$.
The couplings are given in the following table:
\begin{center}
\begin{tabular}{|cc|c|c|c|}\hline
$S_i$ & $S_j$ & $g_{ZS_iS_j}$ & $g_{HWS_i}$ & $g_{WWS_j}$ \\ \hline 
$A^0$ & $h^0$ & $-i \cba / 2\sw\cw$ & $-i / 2\sw$ & $\mw \sba / \sw$\\
$A^0$ & $H^0$ & $i \sba / 2\sw\cw$  & $-i / 2\sw$ & $\mw \cba / \sw$\\ \hline 
\end{tabular}
\end{center}

Fig.~\ref{fig:2HDMdiagrams}.5:
\begin{eqnarray}
{\cal{M}}&=&
-\frac{4\alpha^2}{s-\mvtwo}g_{VS_iS_j}g_{HS_iS_k}g_{WS_jS_k}\left\{
C_{00}(\gver\aone+\gvel\atwo) \right. \nonumber \\
&-&\left.(C_2+C_{12}+C_{22})
\left[\gver(\athree+\afive)+\gvel(\afour+\asix)\right]\right\},
\end{eqnarray}
with the arguments for the integral functions as 
$C(\mwtwo,s,\mhpmtwo,\msktwo,\msjtwo,\msitwo)$.
The couplings are given in the following table:
\begin{center}
\begin{tabular}{|ccc|c|c|c|c|}\hline
$S_i$ & $S_j$ & $S_k$ & $g_{\gamma{S}_iS_j}$
 & $g_{ZS_iS_j}$ & $g_{HS_iS_k}$ & $g_{WS_jS_k}$\\ \hline 
$A^0$ & $h^0$ & $G^{\pm}$
 & 0 & $-i \cba / 2\sw\cw$ & $\ghag$ & $-\sba / 2\sw$\\
$A^0$ & $H^0$ & $G^{\pm}$
 & 0 & $i \sba / 2\sw\cw $ & $\ghag$ & $-\cba / 2\sw$\\
$h^0$ & $A^0$ & $H^{\pm}$
 & 0 & $i \cba / 2\sw\cw$ & $\ghhh$ & $i / 2\sw$\\
$H^0$ & $A^0$ & $H^{\pm}$
 & 0 & $-i \sba / 2\sw\cw$ & $\ghHh$ & $i / 2\sw$\\
$h^0$ & $G^0$ & $G^{\pm}$
 & 0 & $i \sba / 2\sw\cw$ & $\ghhg$ & $i / 2\sw$\\
$H^0$ & $G^0$ & $G^{\pm}$
 & 0 & $i \cba / 2\sw\cw$ & $\ghHg$ & $i / 2\sw$\\
$G^{\pm}$ & $G^{\pm}$ & $h^{0}$
 & 1 & $\cos(2\theta_W) / 2\sw\cw$ & $\ghhg$ & $\sba / 2\sw$\\
$G^{\pm}$ & $G^{\pm}$ & $H^{0}$
 & 1 & $\cos(2\theta_W) / 2\sw\cw$ & $\ghHg$ & $\cba / 2\sw$\\
$H^{\pm}$ & $H^{\pm}$ & $h^{0}$
 & 1 & $\cos(2\theta_W) / 2\sw\cw$ & $\ghhh$ & $\cba / 2\sw$\\
$H^{\pm}$ & $H^{\pm}$ & $H^{0}$
 & 1 & $\cos(2\theta_W) / 2\sw\cw$ & $\ghHh$ & $-\sba / 2\sw$\\ \hline 
\end{tabular}
\end{center}

Fig.~\ref{fig:2HDMdiagrams}.6:
\begin{eqnarray}
{\cal{M}}&=&
\frac{\alpha^2}{s-\mvtwo}g_{VV^{\prime}S_i}g_{WV^{\prime}S_j}g_{HS_iS_j}
C_{0}(\gver\aone+\gvel\atwo),
\end{eqnarray}
with the arguments for the integral functions as 
$C(\mwtwo,s,\mhpmtwo,\msjtwo,m_{V^{\prime}}^2,\msitwo)$.
The couplings are given in the following table:
\begin{center}
\begin{tabular}{|ccc|c|c|c|c|}\hline
$S_i$ & $S_j$ & $V^{\prime}$ & $g_{\gamma{V}^{\prime}S_i}$
 & $g_{Z{V}^{\prime}S_i}$ & $g_{HS_iS_j}$ & $g_{WV^{\prime}S_j}$\\ \hline 
$h^0$ & $G^{\pm}$ & $Z$ & 0 & $\mz \sba / \sw\cw$ & $\ghhg$ & $-\mz\sw$\\
$H^0$ & $G^{\pm}$ & $Z$ & 0 & $\mz \cba / \sw\cw$ & $\ghHg$ & $-\mz\sw$\\
$G^{\pm}$ & $h^{0}$ & $W^{\pm}$ & $\mw$ & $-\mz\sw$ & $\ghhg$ & $\mw\sba/\sw$\\
$G^{\pm}$ & $H^{0}$ & $W^{\pm}$ & $\mw$ & $-\mz\sw$ & $\ghHg$ & $\mw\cba/\sw$\\
\hline
\end{tabular}
\end{center}

Fig.~\ref{fig:2HDMdiagrams}.7:
\begin{eqnarray}
{\cal{M}}&=&
\frac{2\alpha^2}{s-\mvtwo}g_{VV^{\prime}S_i}g_{HV^{\prime}S_j}
g_{WS_iS_j}\left\{
-C_{00}(\gver\aone+\gvel\atwo) \right. \nonumber \\
&&\left. + (-C_2+C_{12}+C_{22})
\left[\gver(\athree+\afive)+\gvel(\afour+\asix)\right]\right\},
\end{eqnarray}
with the arguments for the integral functions as 
$C(\mwtwo,s,\mhpmtwo,\msjtwo,\msitwo,m_{V^{\prime}}^2)$.
The couplings are given in the following table:
\begin{center}
\begin{tabular}{|ccc|c|c|c|c|}\hline
$S_i$&$S_j$&$V^{\prime}$&$g_{\gamma{V}^{\prime}S_i}$
 &$g_{Z{V}^{\prime}S_i}$&$g_{HV^{\prime}S_j}$&$g_{WS_iS_j}$\\ \hline 
$h^0$ & $H^{\pm}$ & $Z$ & 0 & $\mz\sba/\sw\cw$ &
  $\cos(2\theta_W)/2\sw\cw$ & $-\cba/2\sw$\\
$H^0$ & $H^{\pm}$ & $Z$ & 0 & $\mz\cba/\sw\cw$ &
  $\cos(2\theta_W)/2\sw\cw$ & $\sba/2\sw$\\
$G^{\pm}$ & $h^0$ & $W^{\pm}$ & $\mw$ & $-\mz\sw$ &
  $\cba/2\sw$ & $\sba/2\sw$\\
$G^{\pm}$ & $H^0$ & $W^{\pm}$ & $\mw$ & $-\mz\sw$ &
  $-\sba/2\sw$ & $\cba/2\sw$\\
\hline
\end{tabular}
\end{center}

Fig.~\ref{fig:2HDMdiagrams}.8:
\begin{eqnarray}
{\cal{M}}&=&
\frac{\alpha^2}{s-\mztwo}g_{ZZS}g_{HWS}g_{ZWW}
\nonumber \\
&&\left\{
\left[B_0+(2\mhpmtwo+3\mwtwo-2s)C_0-(s-\mhpmtwo-3\mwtwo)C_1
-(s-\mwtwo-3\mhpmtwo)C_2-C_{00}\right](\gver\aone+\gvel\atwo)
\right. \nonumber \\
&+&\left.(2C_0-2C_1+3C_2+C_{12}+C_{22})
\left[\gver(\athree+\afive)+\gvel(\afour+\asix)\right]\right\},
\end{eqnarray}
\end{widetext}
with the arguments for the integral functions as 
$B(s,\mztwo,\mstwo)$, $C(\mwtwo,s,\mhpmtwo,\mwtwo,\mztwo,\mstwo)$.
The couplings are given in the following table:
\begin{center}
\begin{tabular}{|c|c|c|c|}\hline
$S$ & $g_{ZZS}$ & $g_{HWS}$ & $g_{ZWW}$\\ \hline 
$h^0$ & $\mz\sba/\sw\cw$ & $\cba/2\sw$ & $-\cw/\sw$\\
$H^0$ & $\mz\cba/\sw\cw$ & $-\sba/2\sw$ & $-\cw/\sw$\\
\hline
\end{tabular}
\end{center}

Fig.~\ref{fig:2HDMdiagrams}.9:
\begin{eqnarray}
{\cal{M}}&=&
-\frac{\alpha^2 g_{VWHS}g_{WWS}}{s-\mvtwo}B_0(\gver\aone+\gvel\atwo),
\end{eqnarray}
with the arguments for the integral functions as 
$B(\mwtwo, \mstwo,\mwtwo)$.
The couplings are given in the following table:
\begin{center}
\begin{tabular}{|c|c|c|c|}\hline
$S$ & $g_{\gamma{H}WS}$ & $g_{ZHWS}$ & $g_{WWS}$\\ \hline 
$h^0$ & $\cba/2\sw$ & $-\cba/2\cw$ & $\mw\sba/\sw$\\
$H^0$ & $-\sba/2\sw$ & $\sba/2\cw$ & $\mw\cba/\sw$\\
\hline
\end{tabular}
\end{center}

Fig.~\ref{fig:2HDMdiagrams}.10:
\begin{eqnarray}
{\cal{M}}&=&
\frac{\alpha^2 g_{VWS_iS_j}g_{HS_iS_j}}{s-\mvtwo}B_0(\gver\aone+\gvel\atwo),
\end{eqnarray}
with the arguments for the integral functions as 
$B(\mhpmtwo, \msjtwo,\msitwo)$.
The couplings are given in the following table:
\begin{center}
\begin{tabular}{|cc|c|c|c|}\hline
$S_i$ & $S_j$ & $g_{\gamma{W}S_iS_j}$ & $g_{ZWS_iS_j}$ & $g_{HS_iS_j}$\\ 
\hline 
$h^0$ & $H^{\pm}$ & $\cba/2\sw$ & $-\cba/2\cw$ & $\ghhh$\\
$H^0$ & $H^{\pm}$ & $-\sba/2\sw$ & $\sba/2\cw$ & $\ghHh$\\
$h^0$ & $G^{\pm}$ & $\sba/2\sw$ & $-\sba/2\cw$ & $\ghhg$\\
$H^0$ & $G^{\pm}$ & $\cba/2\sw$ & $-\cba/2\cw$ & $\ghHg$\\
\hline
\end{tabular}
\end{center}

Fig.~\ref{fig:2HDMdiagrams}.11:
\begin{eqnarray}
{\cal{M}}&=&
-\frac{\alpha^2 g_{ZHWS}g_{ZZS}}{s-\mztwo}B_0(\gver\aone+\gvel\atwo),
\end{eqnarray}
with the arguments for the integral functions as 
$B(s,\mstwo,\mztwo)$.
The couplings are given in the following table:
\begin{center}
\begin{tabular}{|c|c|c|}\hline
$S$ & $g_{ZHWS}$ & $g_{ZZS}$\\ \hline 
$h^0$ & $-\cba/2\cw$ & $\mz\sba/\sw\cw$\\
$H^0$ & $\sba/2\cw$ & $\mz\cba/\sw\cw$\\
\hline
\end{tabular}
\end{center}

\begin{widetext}

Fig.~\ref{fig:2HDMdiagrams}.12:
\begin{eqnarray}
{\cal{M}}&=&\alpha^2\gw^2g_{WWS}g_{HWS}
\left\{\left[C_0+(\mstwo-u)D_0 
+(\mwtwo-u)D_1+(u-\mhpmtwo)D_3\right]\atwo
-4D_1\asix\right\},
\end{eqnarray}
with the arguments for the integrals as 
$C(\metwo,\metwo,s,\mwtwo,0,\mwtwo)$, 
$D(\mwtwo,\metwo,\metwo,\mhpmtwo,u,s,\mstwo,\mwtwo,0,\mwtwo)$.
The couplings are given in the following table:
\begin{center}
\begin{tabular}{|c|c|c|}\hline
$S$ & $g_{WWS}$ & $g_{HWS}$\\ \hline 
$h^0$ & $\mw\sba/\sw$ & $\cba/2\sw$\\
$H^0$ & $\mw\cba/\sw$ & $-\sba/2\sw$\\
\hline
\end{tabular}
\end{center}

Fig.~\ref{fig:2HDMselfenergy}.1:
\begin{eqnarray}
\Sigma_{W^+H^+}^f(k^2) &=& \frac{N_c\alpha}{2\pi}\gw
\left[(\mt\yl+\mb\yr)B_1+\mb\yr{B}_0\right],
\end{eqnarray}
with the arguments for the integral functions as 
$B(k^2,\mbtwo,\mttwo)$, for $k$ being the external momentum. 

Fig.~\ref{fig:2HDMselfenergy}.2 + \ref{fig:2HDMselfenergy}.3:
\begin{eqnarray}
\Sigma_{W^+H^+}^b(k^2) &=& \frac{\alpha}{8\pi\sw^2}
\left[
\cba\sw\ghhh(2B_1(k^2,\mhtwo,\mhpmtwo)+B_0(k^2,\mhtwo,\mhpmtwo))
\right. \nonumber \\
&-&\left.
\sba\sw\ghHh(2B_1(k^2,\mHtwo,\mhpmtwo)+B_0(k^2,\mHtwo,\mhpmtwo))
\right. \nonumber \\
&+&\left.
\sba\sw\ghhg(2B_1(k^2,\mhtwo,\mwtwo)+B_0(k^2,\mhtwo,\mwtwo))
\right. \nonumber \\
&+&\left.
\cba\sw\ghHg(2B_1(k^2,\mHtwo,\mwtwo)+B_0(k^2,\mHtwo,\mwtwo))
\right. \nonumber \\
&-&\left.
\cba\sba\mw(B_0(k^2,\mhtwo,\mwtwo)-B_1(k^2,\mhtwo,\mwtwo))
\right. \nonumber \\
&+&\left.
\cba\sba\mw(B_0(k^2,\mHtwo,\mwtwo)-B_1(k^2,\mHtwo,\mwtwo))
\right].
\end{eqnarray}

Fig.~\ref{fig:2HDMselfenergy}.4:
\begin{eqnarray}
\Sigma_{G^+H^+}^f(k^2) &=& -\frac{N_c\alpha}{2\pi}
\left\{
(\yl\ygr+\yr\ygl)A(\mttwo)
+\left[
\mb\mt(\yl\ygl+\yr\ygr)+\mb^2(\yl\ygr+\yr\ygl)\right]B_0 \right. \nonumber \\
&+&\left.
k^2(\yl\ygr+\yr\ygl)B_1\right\},
\end{eqnarray}
with the arguments for the integral functions as 
$B(k^2,\mbtwo,\mttwo)$.

Fig.~\ref{fig:2HDMselfenergy}.5 + \ref{fig:2HDMselfenergy}.6:
\begin{eqnarray}
\Sigma_{G^+H^+}^b(k^2) &=& \frac{\alpha}{4\pi}
\left\{
\ghhh\ghhg{B}_0(k^2,\mhtwo,\mhpmtwo)
+\ghHh\ghHg{B}_0(k^2,\mHtwo,\mhpmtwo)
\right. \nonumber \\
&+&\left.
\ghhg\gghg{B}_0(k^2,\mhtwo,\mwtwo)
+\ghHg\ggHg{B}_0(k^2,\mHtwo,\mwtwo)
\right. \nonumber \\
&+&\left.
\frac{\sba\cba}{4\sw^2}
\left[-(k^2+\mhtwo)B_0(k^2,\mhtwo,\mwtwo)
+2k^2{B}_1(k^2,\mhtwo,\mwtwo)\right]
\right. \nonumber \\
&+&\left.
\frac{\sba\cba}{4\sw^2}
\left[(k^2+\mHtwo)B_0(k^2,\mHtwo,\mwtwo)
-2k^2{B}_1(k^2,\mHtwo,\mwtwo)\right]
\right\}
\end{eqnarray}

Fig.~\ref{fig:WHGH}.1:
The real part of the $W^-H^-$ mixing self-energy fixes the 
counterterms for the $\gamma{W}^+{H}^-$ and $Z{W}^+{H}^-$ 
vertices, following Eq.~\ref{eq:counter}:
\begin{eqnarray}
{\cal{M}}&=&\frac{4\pi\alpha}{s-\mvtwo}g_V
{\rm Re}(\Sigma_{W^+H^+}(m^2_{H^{\pm}}))
(\gver\aone+\gvel\atwo),
\end{eqnarray}
where $g_{\gamma}=-1$, $g_Z=\sw/\cw$, and 
$\Sigma_{W^+H^+}=\Sigma_{W^+H^+}^f + \Sigma_{W^+H^+}^b
= {\rm Re}(\Sigma_{W^+H^+})+i {\rm Im}(\Sigma_{W^+H^+})$.

The $W^-H^-$ counterterm cancels the real part of the $W^-H^-$ 
mixing self-energy, following the 
renormalization condition defined in Eq.~\ref{eq:WHrenor}.  
Therefore, only the imaginary part of the $W^-H^-$ mixing self-energy
contributes to Figs.~\ref{fig:WHGH}.2 and \ref{fig:WHGH}.3.  Similarly, 
only the imaginary part of the $G^-H^-$ mixing self-energy
contributes to Fig.~\ref{fig:WHGH}.4; this is related to the imaginary
part of the $W^-H^-$ mixing self-energy by Eq.~\ref{eq:slavnovtaylor2}.

Fig.~\ref{fig:WHGH}.2:
\begin{eqnarray}
{\cal{M}}&=&\frac{4\pi\alpha\gw^2}{\mhpmtwo-\mwtwo}
i{\rm Im}(\Sigma_{W^+H^+}(m^2_{H^{\pm}}))\atwo.
\end{eqnarray}

Fig.~\ref{fig:WHGH}.3:
\begin{eqnarray}
{\cal{M}}&=&\frac{4\pi\alpha{g}_{VWW}}{(s-\mvtwo)(\mhpmtwo-\mwtwo)}(s-\mwtwo)
i{\rm Im}(\Sigma_{W^+H^+}(m^2_{H^{\pm}}))
(\gver\aone+\gvel\atwo),
\end{eqnarray}
where $g_{\gamma{W}W}=-1$ and $g_{ZWW}=-\cw/\sw$.

Fig.~\ref{fig:WHGH}.4:
\begin{eqnarray}
{\cal{M}}&=&-\frac{4\pi\alpha{g}_{VWG} m^2_{H^{\pm}}}
{(s-\mvtwo)(\mhpmtwo-\mwtwo) m_W}
i{\rm Im}(\Sigma_{W^+H^+}(m^2_{H^{\pm}}))(\gver\aone+\gvel\atwo),
\end{eqnarray}
where $g_{\gamma{W}G}=\mw$ and $g_{ZWG}=-\mz\sw$.

\end{widetext}


\section{MSSM contributions}
The matrix elements for $e^+e^- \to W^+H^-$ in the full MSSM are
given here for the first time.  The sfermion triangle that enters 
Fig.~\ref{fig:susydiagrams}.2 was computed in Ref.~\cite{PHI}, 
and agrees with our result.
Fig.~\ref{fig:susydiagrams}.1(b)
with $\chi_i^0$, $\chi_j^0$ and $\chi_k^+$ in the loop 
is analogous to the top/bottom quark triangle diagram and can be
checked by substituting top/bottom quark couplings for the gaugino
couplings; the part involving the $W$ coupling to left-handed gauginos 
is consistent with the top/bottom quark contribution given in the 
previous section.
Formulae were given in Ref.~\cite{ggHW} for the $W^+H^+$
and $G^+H^+$ mixing diagrams,
Fig.~\ref{fig:susyselfenergy}.1-4; our results are consistent
with theirs, although not enough detail was given in Ref.~\cite{ggHW}
to check the overall signs of the diagrams.

We define here our notation for the mixing matrices in the stop/sbottom
and gaugino sectors and various coupling coefficients.

The mixing in the stop sector is defined by:
\begin{equation}
\left(
\begin{array}{c}
\stl\\\str
\end{array}
\right)
=
\left(
\begin{array}{c}
\cat\stone-\sat\sttwo \\
\sat\stone+\cat\sttwo 
\end{array}
\right), 
\end{equation}
where $\tilde t_{L,R}$ are the weak eigenstates and $\tilde t_{1,2}$
are the mass eigenstates, and analogously for the sbottom sector.

The chargino mass matrix is:
\begin{equation} 
X = \left( \begin{array}{cc}
	M_2 & \sqrt{2} m_W \sin\beta \\
	\sqrt{2} m_W \cos\beta & \mu 
	\end{array}
	\right),
\end{equation}
which is diagonalized by the matrices $U$ and $V$ via
$V X^{\dagger} U^{\dagger} = M_D$.

The neutralino mass matrix is:
\begin{widetext}
\begin{equation}
Y = \left( \begin{array}{cccc}
	M_1 & 0 & -m_Z \sw \cos\beta & m_Z \sw \sin\beta \\
	0 & M_2 & m_Z \cw \cos\beta & -m_Z \cw \sin\beta \\
	-m_Z \sw \cos\beta & m_Z \cw \cos\beta & 0 & -\mu \\
	m_Z \sw \sin\beta & -m_Z \cw \sin\beta & -\mu & 0
	\end{array} \right),
\end{equation}
\end{widetext}
which is diagonalized by the matrix $N$ via
$N Y^{\dagger} N^{\dagger} = N_D$.

The matrices that enter the $W^+\chi_i^0\chi_j^-$ couplings are defined as:
\begin{equation}
O_{ij}^L=-\frac{1}{\sqrt{2}}N_{i4}V_{j2}^*+N_{i2}V_{j1}^*,\ \ \ 
O_{ij}^R=\frac{1}{\sqrt{2}}N_{i3}^*U_{j2}+N_{i2}^*U_{j1}.
\end{equation}

The matrices that enter the $Z\chi_i^+\chi_j^-$ couplings are defined as:
\begin{eqnarray}
O_{ij}^{\prime{L}}&=&-V_{i1}V_{j1}^*-\frac{1}{2}V_{i2}V_{j2}^*
+\delta_{ij}\sw^2, \nonumber \\
O_{ij}^{\prime{R}}&=&-U_{i1}^*U_{j1}-\frac{1}{2}U_{i2}^*U_{j2}
+\delta_{ij}\sw^2.
\end{eqnarray}

The matrices that enter the $Z\chi_i^0\chi_j^0$ couplings are defined as:
\begin{equation}
O_{ij}^{\prime\prime{L}} = -\frac{1}{2}N_{i3}N_{j3}^*
+\frac{1}{2}N_{i4}N_{j4}^*, \ \ \ 
O_{ij}^{\prime\prime{R}} = -O_{ij}^{\prime\prime{L}*}.
\end{equation}

The matrices that enter the $H^{-}\chi_i^0\chi_j^+$ couplings are defined as:
\begin{eqnarray}
Q_{ij}^{\prime{L}}&=&N_{i4}^*V_{j1}^*
+\frac{1}{\sqrt{2}}(N_{i2}^*+N_{i1}^*\tan\theta_W)V_{j2}^*, \nonumber \\
Q_{ij}^{\prime{R}}&=&N_{i3}U_{j1}-\frac{1}{\sqrt{2}}
(N_{i2}+N_{i1}\tan\theta_W)U_{j2}.
\end{eqnarray}

The $H^- \tilde f \tilde f^{\prime}$ coupling coefficients are given 
in the following table:
\begin{widetext}
\begin{center}
\begin{tabular}{|c|c|} \hline
$\ghenu$&$-\mw\sin{2}\beta/\sqrt{2}\sw$ \\
$\ghdu$&$-\mw\sin{2}\beta/\sqrt{2}\sw$ \\
$g_{H^-\sbl\stl}$&$-\mw\sin{2}\beta/\sqrt{2}\sw 
         + (\mb^2\tan\beta + \mt^2\cot\beta)/\sqrt{2}\mw\sw$ \\
$g_{H^-\sbl\str}$&$\mt(\mu + A_t\cot\beta)/\sqrt{2}\mw\sw$ \\
$g_{H^-\sbr\stl}$&$\mb(\mu + A_b\tan\beta)/\sqrt{2}\mw\sw$\\
$g_{H^-\sbr\str}$&$\mt\mb(\tan\beta+\cot\beta)/\sqrt{2}\mw\sw$\\ 
\hline
$\ghbonetone$&$\cat\cab{g}_{H^-\sbl\stl}+\sat\cab{g}_{H^-\sbl\str}+\cat\sab{g}_{H^-\sbr\stl}+\sat\sab{g}_{H^-\sbr\str}$ \\
$\ghbonettwo$&$-\sat\cab{g}_{H^-\sbl\stl}+\cat\cab{g}_{H^-\sbl\str}-\sat\sab{g}_{H^-\sbr\stl}+\cat\sab{g}_{H^-\sbr\str}$ \\
$\ghbtwotone$&$-\cat\sab{g}_{H^-\sbl\stl}-\sat\sab{g}_{H^-\sbl\str}+\cat\cab{g}_{H^-\sbr\stl}+\sat\cab{g}_{H^-\sbr\str}$ \\
$\ghbtwottwo$&$\sat\sab{g}_{H^-\sbl\stl}-\cat\sab{g}_{H^-\sbl\str}-\sat\cab{g}_{H^-\sbr\stl}+\cat\cab{g}_{H^-\sbr\str}$ \\
 \hline
\end{tabular}
\end{center}

The $G^- \tilde f \tilde f^{\prime}$ coupling coefficients are given 
in the following table:
\begin{center}
\begin{tabular}{|c|c|} \hline
$\ggenu$&$\mw\cos{2}\beta/\sqrt{2}\sw$  \\
$\ggdu$&$\mw\cos{2}\beta/\sqrt{2}\sw$ \\ 
$g_{G^-\sbl\stl}$&$\mw\cos{2}\beta/\sqrt{2}\sw 
        + (\mt^2-\mb^2)/\sqrt{2}\mw\sw$ \\
$g_{G^-\sbl\str}$&$\mt(-\mu\cot\beta+A_t)/\sqrt{2}\mw\sw$ \\
$g_{G^-\sbr\stl}$&$-\mb(-\mu\tan\beta+A_b)/\sqrt{2}\mw\sw$ \\
$g_{G^-\sbr\str}$&0 \\ \hline 
$\ggbonetone$&$\cat\cab{g}_{G^-\sbl\stl}+\sat\cab{g}_{G^-\sbl\str}+\cat\sab{g}_{G^-\sbr\stl}$ \\
$\ggbonettwo$&$-\sat\cab{g}_{G^-\sbl\stl}+\cat\cab{g}_{G^-\sbl\str}-\sat\sab{g}_{G^-\sbr\stl}$ \\
$\ggbtwotone$&$-\cat\sab{g}_{G^-\sbl\stl}-\sat\sab{g}_{G^-\sbl\str}+\cat\cab{g}_{G^-\sbr\stl}$ \\
$\ggbtwottwo$&$\sat\sab{g}_{G^-\sbl\stl}-\cat\sab{g}_{G^-\sbl\str}-\sat\cab{g}_{G^-\sbr\stl}$ \\
 \hline
\end{tabular}
\end{center}

We now list our results for the MSSM diagrams.  It is to be understood
that diagrams involving charginos $\chi^+_i$ are summed over $i=1,2$
and diagrams involving neutralinos $\chi^0_i$ are summed over
$i=1,\ldots,4$.

Fig.~\ref{fig:susydiagrams}.1(a):
\begin{eqnarray}
{\cal{M}}&=&-\frac{2\alpha^2}{s-\mvtwo} \nonumber \\
&&
\left\{\left[
(F-G+H)B_0+
(\mktwo(F-G+H)-k_1\cdot{k}_2H-\mi\mj\mk{J})C_0
+(\mwtwo{F}+k_1\cdot{k}_2G-(k_1\cdot{k}_2-\mwtwo)H)C_1
\right. \right. \nonumber \\
&+& \left. \left.
(-k_1\cdot{k}_2F-\mhpmtwo{G}-(k_1\cdot{k}_2-\mhpmtwo)H)C_2
-2(F+H)C_{00}\right](\gver\aone+\gvel\atwo)
\right.  \nonumber \\
&+&\left.
\left[
HC_0+(-G+H)C_1+(F+3H)C_2+2(F+H)(C_{12}+C_{22})\right]
\left[\gver(\athree+\afive)+\gvel(\afour+\asix)\right]
\right. \nonumber \\
&+& \left.
\left[H_{im}C_0+(G_{im}+H_{im})C_1+(F_{im}+H_{im})C_2\right]
(\gver\aseven+\gvel\aeight)\right\}
\end{eqnarray}
where 
\begin{eqnarray}
&&F=\mi(\ghl\gvl\gwl+\ghr\gvr\gwr),\ \ \ G=\mj(\ghl\gvr\gwl+\ghr\gvl\gwr),
\ \ \ H=\mk(\ghr\gvl\gwl+\ghl\gvr\gwr),\nonumber \\
&&J=\ghr\gvr\gwl+\ghl\gvl\gwr,
\nonumber \\
&&F_{im}=\mi(\ghl\gvl\gwl-\ghr\gvr\gwr),\ 
G_{im}=\mj(\ghl\gvr\gwl-\ghr\gvl\gwr),
\ H_{im}=\mk(\ghr\gvl\gwl-\ghl\gvr\gwr).
\end{eqnarray}
The arguments for the integral functions are
$B(s,\mjtwo,\mitwo)$, $C(\mwtwo,s,\mhpmtwo,\mktwo,\mjtwo,\mitwo)$.

Fig.~\ref{fig:susydiagrams}.1(b): 
Similar to Fig.~\ref{fig:susydiagrams}.1(a), under the exchange of 
\begin{eqnarray}
\ghl\leftrightarrow\ghr, \ \ \ \aseven, \aeight\ {\rm terms\ flip\ sign}. 
\end{eqnarray}
The couplings are given in the following table:
\begin{center}
\begin{tabular}{|c|ccc|c|c|c|c|c|c|c|c|}\hline
&$\chi_i$&$\chi_j$&$\chi_k$&
$\gpl$&$\gpr$&$\gzl$&$\gzr$&$\ghl$&$\ghr$&$\gwl$&$\gwr$ \\  \hline
\ref{fig:susydiagrams}.1(a)&$\chi_i^+$&$\chi_j^+$&$\chi_k^0$&
$-\delta_{ij}$&$-\delta_{ij}$&$O_{ji}^{\prime{L}}/\sw\cw$
&$O_{ji}^{\prime{R}}/\sw\cw$
&$-Q_{ki}^{\prime{R}*}\sinb/\sw$
&$-Q_{ki}^{\prime{L}*}\cosb/\sw$
&$O_{kj}^{L}/\sw$
&$O_{kj}^{R}/\sw$  \\
\ref{fig:susydiagrams}.1(b)&$\chi_i^0$&$\chi_j^0$&$\chi_k^+$&
0&0&$O_{ij}^{\prime\prime{L}}/\sw\cw$
&$O_{ij}^{\prime\prime{R}}/\sw\cw$
&$-Q_{ik}^{\prime{R}*}\sinb/\sw$
&$-Q_{ik}^{\prime{L}*}\cosb/\sw$
&$O_{jk}^{L}/\sw$
&$O_{jk}^{R}/\sw$  \\
\hline
\end{tabular}
\end{center}

Fig.~\ref{fig:susydiagrams}.2:
\begin{eqnarray}
{\cal{M}}&=&
-\frac{4N_c\alpha^2}{s-\mvtwo}g_{VS_iS_j}g_{HS_iS_k}g_{WS_jS_k}\left\{
C_{00}(\gver\aone+\gvel\atwo) \right. \nonumber \\
&-&\left.(C_2+C_{12}+C_{22})
\left[\gver(\athree+\afive)+\gvel(\afour+\asix)\right]\right\},
\end{eqnarray}
with the arguments for the integral functions as 
$C(\mwtwo,s,\mhpmtwo,\msktwo,\msjtwo,\msitwo)$, $N_c=1$ for sleptons and
$N_c=3$ for squarks. 
The couplings are given in the following table:
\begin{center}
\begin{tabular}{|ccc|c|c|c|c|}\hline
$S_i$&$S_j$&$S_k$&$g_{\gamma{S}_iS_j}$
&$g_{ZS_iS_j}$&$g_{HS_iS_k}$&$g_{WS_jS_k}$\\ \hline 
$\snu$&$\snu$&$\sel$&$-\gpnul$&$-\gznul$&$\ghenu$&$-\gw$  \\
$\sel$&$\sel$&$\snu$&$\gpel$&$\gzel$&$\ghenu$&$\gw$  \\
$\sul$&$\sul$&$\sdl$&$-\gpul$&$-\gzul$&$\ghdu$&$-\gw$  \\
$\sdl$&$\sdl$&$\sul$&$\gpdl$&$\gzdl$&$\ghdu$&$\gw$  \\
$\tilde{t}_i$&$\tilde{t}_j$&$\tilde{b}_k$&
$-\gpul{M}^t_{Li}{M}^t_{Lj}-\gpur{M}^t_{Ri}{M}^t_{Rj}$&
$-\gzul{M}^t_{Li}{M}^t_{Lj}-\gzur{M}^t_{Ri}{M}^t_{Rj}$&
$g_{H\tilde{b}_k\tilde{t}_i}$&$-\gw{M}^t_{Lj}{M}^b_{Lk}$ \\
$\tilde{b}_i$&$\tilde{b}_j$&$\tilde{t}_k$&
$\gpdl{M}^b_{Li}{M}^b_{Lj}+\gpdr{M}^b_{Ri}{M}^b_{Rj}$&
$\gzdl{M}^b_{Li}{M}^b_{Lj}+\gzdr{M}^b_{Ri}{M}^b_{Rj}$&
$g_{H\tilde{b}_i\tilde{t}_k}$&$\gw{M}^t_{Lk}{M}^b_{Lj}$ \\ \hline
\end{tabular}
\end{center}
where 
$M^{t}_{L1}=M^{t}_{R2}=\cat$, $-M^{t}_{L2}=M^{t}_{R1}=\sat$, 
$M^{b}_{L1}=M^{b}_{R2}=\cab$, and $-M^{b}_{L2}=M^{b}_{R1}=\sab$.

Fig.~\ref{fig:susydiagrams}.3:
\begin{eqnarray}
{\cal{M}}&=&\alpha^2\gw\gchiiel\gchijnur\left[(C_0+C_2)\mi\ghl+C_2\mj\ghr\right]\atwo,
\end{eqnarray}
with the arguments for the integral functions as 
$C(0,u,\mhpmtwo,\mitwo,\msltwo,\mjtwo)$.
The couplings are given in the following table:
\begin{center}
\begin{tabular}{|ccc|c|c|c|c|}\hline
$\chi_i$&$\chi_j$&$\slepton$&$\ghl$&$\ghr$&$\gchiiel$&$\gchijnur$ \\ \hline
$\chi_i^0$&$\chi_j^+$&$\sel$
&$-Q_{ij}^{\prime{R}*}\sinb/\sw$
&$-Q_{ij}^{\prime{L}*}\cosb/\sw$
&$N_{i2}^*/\sqrt{2}\sw + N_{i1}^*/\sqrt{2}\cw$
&$-U_{j1}/\sw$ \\
$\chi_i^+$&$\chi_j^0$&$\snu$
&$-Q_{ji}^{\prime{R}*}\sinb/\sw$
&$-Q_{ji}^{\prime{L}*}\cosb/\sw$
&$-V_{i1}^*/\sw$
&$-N_{j2}/\sqrt{2}\sw + N_{j1}/\sqrt{2}\cw$ \\ \hline
\end{tabular}
\end{center}

Fig.~\ref{fig:susydiagrams}.4:
\begin{eqnarray}
{\cal{M}}&=&-\alpha^2\gw\gchiiesel\gchiinusnur\ghenu{C}_2\atwo,
\end{eqnarray}
with the arguments for the integral functions as 
$C(0,u,\mhpmtwo,\mseltwo,\mitwo,\msnutwo)$.
The couplings are given in the following table:
\begin{center}
\begin{tabular}{|c|c|c|}\hline
$\chi_i$&$\gchiiesel$&$\gchiinusnur$\\ \hline
$\chi_i^0$
&$N_{i2}^*/\sqrt{2}\sw + N_{i1}^*/\sqrt{2}\cw$
&$-N_{i2}/\sqrt{2}\sw + N_{i1}/\sqrt{2}\cw$ \\ \hline
\end{tabular}
\end{center}

Fig.~\ref{fig:susydiagrams}.5(a,c):
\begin{eqnarray}
{\cal{M}}&=&\alpha^2\gchiie\gchike
\left\{
\left[
\mi\gwr\ghl\left(-C_0-\mjtwo{D}_0-\mwtwo{D}_1-u(D_2+D_3)+2D_{00}\right)
\right.\right. \nonumber \\
&+&\left.\left.
\mj\gwr\ghr\left(-C_0-
(u+\mjtwo)D_0-(u+\mwtwo)D_1-2uD_2-(u+\mhpmtwo)D_3+2D_{00}\right)
\right.\right. \nonumber \\
&+&\left.\left.
\mk\gwl\ghr\left(C_0+\mjtwo{D}_0+u(D_1+D_2)+\mhpmtwo{D}_3\right)
+\mi\mj\mk\gwl\ghl{D}_0 \right]\atwo
\right. \nonumber \\
&+&\left.
\left[
\mi\gwr\ghl\left(-2(D_{33}+D_{13}+D_{23}\right)+
\mj\gwr\ghr\left(-2D_3-2(D_{33}+D_{13}+D_{23})\right)\right]\afour
\right. \nonumber \\
&+&\left.
\left[
\mi\gwr\ghl\left(-2(D_2+D_3)-2(D_{22}+D_{33}+D_{12}+D_{13}+D_{23})\right)
\right. \right. \nonumber \\
&+& \left. \left.
\mj\gwr\ghr\left(-2D_0-(2D_1+4D_2+4D_3)-2(D_{22}+D_{33}+D_{12}+D_{13}+2D_{23})
\right)\right. \right. \nonumber \\
&+&\left. \left. 
\mk\gwl\ghr{2}D_1 \right]\asix \right\},
\end{eqnarray}
with the arguments for the integral functions as 
$C(0,0,s,\mktwo,\msltwo,\mitwo)$,
$D(\mwtwo,0,0,\mhpmtwo,u,s,\mjtwo,\mktwo,\msltwo,\mitwo)$.
The couplings are given in the following table:
\begin{center}
\begin{tabular}{|c|cccc|c|c|c|c|c|c|}\hline
&$\chi_i$&$\chi_j$&$\chi_k$&$\slepton$ 
  &$\gchiie$&$\gchike$&$\ghl$&$\ghr$&$\gwl$&$\gwr$ \\ \hline
\ref{fig:susydiagrams}.5(a)&$\chi_i^0$&$\chi_j^+$&$\chi_k^0$&$\sel$&
$\frac{N_{i2}^{*}}{\sqrt{2}\sw} + \frac{N_{i1}^{*}}{\sqrt{2}\cw}$ &
$\frac{N_{k2}}{\sqrt{2}\sw} + \frac{N_{k1}}{\sqrt{2}\cw}$ &
$-Q_{ij}^{\prime{R}*}\sinb/\sw$ &
$-Q_{ij}^{\prime{L}*}\cosb/\sw$ &
$O_{kj}^L/\sw$ & $O_{kj}^R/\sw$  \\

\ref{fig:susydiagrams}.5(b)&$\chi_i^0$&$\chi_j^+$&$\chi_k^0$&$\ser$&
$-\sqrt{2}N_{i1}/\cw$ & $-\sqrt{2}N_{k1}^*/\cw$ &
$-Q_{ij}^{\prime{R}*}\sinb/\sw$ &
$-Q_{ij}^{\prime{L}*}\cosb/\sw$ &
$O_{kj}^L/\sw$ & $O_{kj}^R/\sw$  \\

\ref{fig:susydiagrams}.5(c)&$\chi_i^+$&$\chi_j^0$&$\chi_k^+$&$\snu$&
$-V_{i1}^*/\sw$ & $-V_{k1}/\sw$ &
$-Q_{ji}^{\prime{R}*}\sinb/\sw$ &
$-Q_{ji}^{\prime{L}*}\cosb/\sw$ &
$-O_{jk}^R/\sw$ & $-O_{jk}^L/\sw$  \\

\ref{fig:susydiagrams}.6(a)&$\chi_i^0$&$\chi_j^+$&$\chi_k^0$&$\sel$&
$\frac{N_{i2}^{*}}{\sqrt{2}\sw}+\frac{N_{i1}^{*}}{\sqrt{2}\cw}$&
$\frac{N_{k2}}{\sqrt{2}\sw}+\frac{N_{k1}}{\sqrt{2}\cw}$&
$-Q_{kj}^{\prime{R}*}\sinb/\sw$ &
$-Q_{kj}^{\prime{L}*}\cosb/\sw$ &
$-O_{ij}^R/\sw$ & $-O_{ij}^L/\sw$  \\

\ref{fig:susydiagrams}.6(b)&$\chi_i^0$&$\chi_j^+$&$\chi_k^0$&$\ser$&
$-\sqrt{2}N_{i1}/\cw$ & $-\sqrt{2}N_{k1}^*/\cw$ &
$-Q_{kj}^{\prime{R}*}\sinb/\sw$ &
$-Q_{kj}^{\prime{L}*}\cosb/\sw$ &
$-O_{ij}^R/\sw$ & $-O_{ij}^L/\sw$  \\  \hline

\end{tabular}
\end{center}

Fig.~\ref{fig:susydiagrams}.5(b):
Similar to Fig.~\ref{fig:susydiagrams}.5(a,c), under the exchange of 
\begin{equation}
\ghl\leftrightarrow\ghr,\ \ \ \gwl\leftrightarrow\gwr,
\ \ \ \atwo\leftrightarrow\aone,\ \ \ \afour\leftrightarrow\athree,
\ \ \ \asix\leftrightarrow\afive,
\end{equation}

Fig.~\ref{fig:susydiagrams}.6(a):
\begin{eqnarray}
{\cal{M}}&=&\alpha^2\gchiie\gchije
\left\{
\left[
\mi\gwl\ghl\left(C_0+\mjtwo{D}_0+\mhpmtwo{D}_1+t(D_2+D_3)\right)
\right.\right. \nonumber \\
&+&\left.\left.
\mj\gwr\ghl\left(-C_0-
(\mjtwo+t)D_0-(t+\mhpmtwo)D_1-2tD_2-(t+\mwtwo)D_3+2D_{00}\right)
\right.\right. \nonumber \\
&+&\left.\left.
\mk\gwr\ghr\left(-C_0-\mjtwo{D}_0-t(D_1+D_2)-\mwtwo{D}_3+2D_{00}\right)
+\mi\mj\mk\gwl\ghr{D}_0 \right]\atwo
\right. \nonumber \\
&+&\left.
\left[
\mi\gwl\ghl{2}D_3+
\mj\gwr\ghl\left(-2D_0-(4D_1+4D_2+2D_3)-2(D_{11}+D_{22}+2D_{12}+D_{13}+D_{23})\right)
\right. \right. \nonumber \\
&+&\left. \left.
\mk\gwr\ghr
\left(-2(D_1+D_2)-2(D_{11}+D_{22}+2D_{12}+D_{13}+D_{23})\right)
\right]\afour
\right. \nonumber \\
&+&\left.
\left[
\mj\gwr\ghl\left(-2D_1-2(D_{11}+D_{12}+D_{13})\right)
+\mk\gwr\ghr(-2)(D_{11}+D_{12}+D_{13}) \right]\asix \right\},
\end{eqnarray}
with the arguments for the integral functions as 
$C(0,0,s,\mktwo,\msltwo,\mitwo)$,
$D(\mhpmtwo,0,0,\mwtwo,t,s,\mjtwo,\mktwo,\msltwo,\mitwo)$.

Fig.~\ref{fig:susydiagrams}.6(b):
Similar to Fig.~\ref{fig:susydiagrams}.6(a), under the exchange of 
\begin{equation}
\ghl\leftrightarrow\ghr,\ \ \ \gwl\leftrightarrow\gwr,
\ \ \ \atwo\leftrightarrow\aone,\ \ \ \afour\leftrightarrow\athree,
\ \ \ \asix\leftrightarrow\afive,
\end{equation}

Fig.~\ref{fig:susydiagrams}.7:
\begin{eqnarray}
{\cal{M}}&=&-2\alpha^2\gchiiesel\gchiieser\ghenu\gw
\left[D_{00}\atwo-(D_3+D_{33}+D_{13}+D_{23})\afour
\right. \nonumber \\
&-&\left.
(D_{2}+D_3+D_{22}+D_{33}+D_{12}+D_{13}+2D_{23})\asix \right],
\end{eqnarray}
with the arguments for the integral functions as 
$D(\mwtwo,0,0,\mhpmtwo,u,s,\msnutwo,\mseltwo,\mitwo,\mseltwo)$.
The couplings are given in the following table:
\begin{center}
\begin{tabular}{|c|c|c|}\hline
$\chi_i$&$\gchiiesel$&$\gchiieser$ \\ \hline
$\chi_i^0$ & $N_{i2}^*/\sqrt{2}\sw + N_{i1}^*/\sqrt{2}\cw$
& $N_{i2}/\sqrt{2}\sw + N_{i1}/\sqrt{2}\cw$\\ \hline
\end{tabular}
\end{center}

Fig.~\ref{fig:susydiagrams}.8:
\begin{eqnarray}
{\cal{M}}&=&-2\alpha^2\gchiiesnul\gchiiesnur\ghenu\gw
\left[D_{00}\atwo-(D_1+D_2+D_{11}+D_{22}+2D_{12}+D_{13}+D_{23})\afour
\right. \nonumber \\
&-&\left.
(D_1+D_{11}+D_{12}+D_{13})\asix \right],
\end{eqnarray}
where the arguments of the integral functions are
$D(\mhpmtwo,0,0,\mwtwo,t,s,\mseltwo,\msnutwo,\mitwo,\msnutwo)$.
The couplings are given in the following table:
\begin{center}
\begin{tabular}{|c|c|c|}\hline
$\chi_i$&$\gchiiesnul$&$\gchiiesnur$ \\ \hline
$\chi_i^+$ & $-V_{i1}^*/\sw$
&$-V_{i1}/\sw$ \\ \hline
\end{tabular}
\end{center}

Fig.~\ref{fig:susydiagrams}.9:
\begin{eqnarray}
{\cal{M}}&=&-2\alpha^2\gchiiesnul\gchijeser\gw \nonumber \\
&&\left\{
D_{00}(\mi\ghl+\mj\ghr)\atwo
+\left[(D_3+D_{13}+D_{23})\mi\ghl+(D_{13}+D_{23})\mj\ghr\right]\afour
\right. \nonumber \\
&+&\left.
\left[-(D_2+D_{12}+D_{22})\mi\ghl-(D_{12}+D_{22})\mj\ghr\right]\asix \right\},
\end{eqnarray}
with the arguments for the integral functions as 
$D(\mwtwo,0,\mhpmtwo,0,u,t,\msnutwo,\mseltwo,\mjtwo,\mitwo)$.
The couplings are given in the following table:
\begin{center}
\begin{tabular}{|cc|c|c|c|c|c|}\hline
$\chi_i$&$\chi_j$&$\ghl$&$\ghr$&$\gchiiesnul$&$\gchijeser$  \\ \hline
$\chi_i^+$&$\chi_j^0$ & $-Q_{ji}^{\prime{R}*}\sinb/\sw$ &
$-Q_{ji}^{\prime{L}*}\cosb/\sw$
& $-V_{i1}^*/\sw$
& $N_{j2}/\sqrt{2}\sw + N_{j1}/\sqrt{2}\cw$
\\ \hline
\end{tabular}
\end{center}

Fig.~\ref{fig:susydiagrams}.10:
\begin{eqnarray}
{\cal{M}}&=&\alpha^2\gchiiesnur\gchijesel\ghenu \nonumber \\
&&\left\{
\left[\mi\mj\gwl{D}_{0}+\gwr(-C_0-\mjtwo{D}_0-\mwtwo{D}_1-u{D}_2+2D_{00})\right]\atwo\right. \nonumber \\
&+&\left.
2\gwr(D_{13}+D_{23})\afour
-2\gwr(D_2+D_{12}+D_{22})\asix \right\},
\end{eqnarray}
with the arguments for the integrals as 
$C(0,\mhpmtwo,t,\mitwo,\msnutwo,\mseltwo)$,
$D(\mwtwo,0,\mhpmtwo,0,u,t,\mjtwo,\mitwo,\msnutwo,\mseltwo)$.
The couplings are given in the following table:
\begin{center}
\begin{tabular}{|cc|c|c|c|c|}\hline
$\chi_i$&$\chi_j$&$\gwl$&$\gwr$&$\gchiiesnur$&$\gchijesel$ \\ \hline
$\chi_i^+$&$\chi_j^0$ & $-O_{ji}^{R}/\sw$ &
$-O_{ji}^{L}/\sw$
& $-V_{i1}/\sw$
& $N_{j2}^*/\sqrt{2}\sw + N_{j1}^*/\sqrt{2}\cw$
\\ \hline
\end{tabular}
\end{center}

Fig.~\ref{fig:susydiagrams}.11:
\begin{eqnarray}
{\cal{M}}&=&\frac{N_c\alpha^2}{s-\mvtwo}g_{VWS_iS_j}g_{HS_iS_j}B_0
(\gver\aone+\gvel\atwo),
\end{eqnarray}
with the arguments for the integral functions as 
$B(\mhpmtwo,\msjtwo,\msitwo)$, $N_c=1$ for sleptons and $N_c=3$ for squarks.
The couplings are given in the following table:
\begin{center}
\begin{tabular}{|cc|c|c|c|} \hline
$S_i$&$S_j$&$g_{HS_iS_j}$&$g_{\gamma{W}S_iS_j}$&$g_{ZWS_iS_j}$ \\ \hline
$\snu$ & $\sel$ & $\ghenu$ & $(\ee+\enu)/\sqrt{2}\sw$
 & $-(\ee+\enu)/\sqrt{2}\cw$  \\
$\sul$ & $\sdl$ & $\ghdu$ & $(\eu+\ed)/\sqrt{2}\sw$
 & $-(\eu+\ed)/\sqrt{2}\cw$  \\
$\stone$ & $\sbone$ & $\ghbonetone$ & $(\eu+\ed)\cat\cab/\sqrt{2}\sw$
 & $-(\eu+\ed)\cat\cab/\sqrt{2}\cw$  \\
$\stone$ & $\sbtwo$ & $\ghbtwotone$ & $-(\eu+\ed)\cat\sab/\sqrt{2}\sw$
 & $(\eu+\ed)\cat\sab/\sqrt{2}\cw$  \\
$\sttwo$ & $\sbone$ & $\ghbonettwo$ & $-(\eu+\ed)\sat\cab/\sqrt{2}\sw$
 & $(\eu+\ed)\sat\cab/\sqrt{2}\cw$  \\ 
$\sttwo$ & $\sbtwo$ & $\ghbtwottwo$ & $(\eu+\ed)\sat\sab/\sqrt{2}\sw$
 & $-(\eu+\ed)\sat\sab/\sqrt{2}\cw$  \\  \hline
\end{tabular}
\end{center}

Fig.~\ref{fig:susyselfenergy}.1:
\begin{eqnarray}
\Sigma_{W^+H^+}^f(k^2) &=&\frac{\alpha}{2\pi}\left\{
\left[\mj(\ghl\gwl+\ghr\gwr)+\mi(\ghl\gwr+\ghr\gwl)\right]B_1
+\mi(\ghl\gwr+\ghr\gwl)B_0\right\},
\end{eqnarray}
with the arguments for the integral functions as 
$B(k^2,\mitwo,\mjtwo)$, for $k$ being the external momentum. 
The couplings are given in the following table:
\begin{center}
\begin{tabular}{|cc|c|c|c|c|} \hline
$\chi_i$&$\chi_j$&$\ghl$&$\ghr$&$\gwl$&$\gwr$ \\ \hline 
$\chi_i^0$ & $\chi_j^+$ & $-Q_{ij}^{\prime{R}*}\sinb/\sw$ &
$-Q_{ij}^{\prime{L}*}\cosb/\sw$ &
$O_{ij}^L/\sw$ & $O_{ij}^R/\sw$ \\ \hline
\end{tabular}
\end{center}

Fig.~\ref{fig:susyselfenergy}.2:
\begin{eqnarray}
\Sigma_{W^+H^+}^b(k^2) 
&=&-\frac{N_c\alpha}{4\pi}g_{HS_iS_j}g_{WS_iS_j}(2B_1+B_0),
\end{eqnarray}
with the arguments for the integral functions as 
$B(k^2,\msjtwo,\msitwo)$, for $k$ being the external momentum. 
The couplings are given in the following table:
\begin{center}
\begin{tabular}{|cc|c|c|} \hline
$S_i$&$S_j$&$g_{HS_iS_j}$&$g_{WS_iS_j}$ \\ \hline
$\snu$&$\sel$&$\ghenu$&$\gw$ \\
$\sul$&$\sdl$&$\ghdu$&$\gw$  \\
$\stone$&$\sbone$&$\ghbonetone$&$\gw\cat\cab$  \\
$\stone$&$\sbtwo$&$\ghbtwotone$&$-\gw\cat\sab$  \\
$\sttwo$&$\sbone$&$\ghbonettwo$&$-\gw\sat\cab$  \\ 
$\sttwo$&$\sbtwo$&$\ghbtwottwo$&$\gw\sat\sab$  \\  \hline
\end{tabular}
\end{center}

Fig.~\ref{fig:susyselfenergy}.3:
\begin{eqnarray}
\Sigma_{G^+H^+}^f(k^2) 
&=&-\frac{\alpha}{2\pi}\left\{(\ghl\ggr+\ghr\ggl)A(\mjtwo)+
\left[\mitwo(\ghl\ggr+\ghr\ggl)+\mi\mj(\ghl\ggl+\ghr\ggr)\right]B_0 \right.
\nonumber \\
&+&\left.
k^2(\ghl\ggr+\ghr\ggl)B_1\right\},
\end{eqnarray}
with the arguments for the integral functions as 
$B(k^2,\mitwo,\mjtwo)$, for $k$ being the external momentum. 
The couplings are given in the following table:
\begin{center}
\begin{tabular}{|cc|c|c|c|c|} \hline
$\chi_i$&$\chi_j$&$\ghl$&$\ghr$&$\ggl$&$\ggr$ \\ \hline 
$\chi_i^0$ & $\chi_j^+$ & $-Q_{ij}^{\prime{R}*}\sinb/\sw$ &
$-Q_{ij}^{\prime{L}*}\cosb/\sw$ &
$-Q_{ij}^{\prime{L}}\sinb/\sw$ &
$Q_{ij}^{\prime{R}}\cosb/\sw$ \\ \hline
\end{tabular}
\end{center}

Fig.~\ref{fig:susyselfenergy}.4:
\begin{eqnarray}
\Sigma_{G^+H^+}^b(k^2) &=&\frac{N_c\alpha}{4\pi}g_{HS_iS_j}g_{GS_iS_j}B_0,
\end{eqnarray}
with the arguments for the integral functions as 
$B(k^2,\msjtwo,\msitwo)$, for $k$ being the external momentum. 
The couplings are given in the following table:
\begin{center}
\begin{tabular}{|cc|c|c|} \hline
$S_i$&$S_j$&$g_{HS_iS_j}$&$g_{GS_iS_j}$ \\ \hline
$\snu$&$\sel$&$\ghenu$&$\ggenu$ \\
$\sul$&$\sdl$&$\ghdu$&$\ggdu$ \\
$\stone$&$\sbone$&$\ghbonetone$&$\ggbonetone$ \\
$\stone$&$\sbtwo$&$\ghbtwotone$&$\ggbtwotone$ \\
$\sttwo$&$\sbone$&$\ghbonettwo$&$\ggbonettwo$ \\
$\sttwo$&$\sbtwo$&$\ghbtwottwo$&$\ggbtwottwo$ \\  \hline
\end{tabular}
\end{center}

\end{widetext}


\end{document}